\documentstyle[emulateapj,flushrt,tighten]{article} 
\input{psfig.sty}
\singlespace
\newcommand{\kms}{\,{\rm km\,s^{-1}}}
\newcommand{\msun}{\,{\rm M_\odot}}
\newcommand{\mden}{\,{\rm M_\odot\,Mpc^{-3}}}
\newcommand{\etal}{{et al.\ }}
\newcommand{\vir}{{\rm vir}}
\newcommand{\DM}{{\rm DM}}
\newcommand{\intd}{{\rm int}}
\newcommand{\ej}{{\rm ej}}
\newcommand{\beq}{\begin{equation}}
\newcommand{\eeq}{\end{equation}}
\newcommand{\ba}{\begin{eqnarray}}
\newcommand{\ea}{\end{eqnarray}}
\def\spose#1{\hbox to 0pt{#1\hss}}
\newcommand{\lta}{\mathrel{\spose{\lower 3pt\hbox{$\mathchar"218$}}
      \raise 2.0pt\hbox{$\mathchar"13C$}}}
\newcommand{\gta}{\mathrel{\spose{\lower 3pt\hbox{$\mathchar"218$}}
      \raise 2.0pt\hbox{$\mathchar"13E$}}}

\lefthead{Volonteri, Haardt, \& Madau}
\righthead{Assembly and Merger History of SMBHs}
\submitted{ApJ, in press}
\makeatletter

\newenvironment{figurehere}
  {\def\@captype{figure}}
  {}
\makeatother

\begin{document}

\title{The Assembly and Merging History of Supermassive Black Holes in Hierarchical 
Models of Galaxy Formation}

\author{Marta Volonteri\altaffilmark{1,2}, Francesco Haardt\altaffilmark{2}, \& 
Piero Madau\altaffilmark{3}}

\altaffiltext{1}{Dipartimento di Fisica, Universit\`a di Milano Bicocca, Italy.}
\altaffiltext{2}{Dipartimento di Scienze, Universit\`a dell'Insubria/Sede di Como, 
Italy.} 
\altaffiltext{3}{Department of Astronomy \& Astrophysics, University of California, 
Santa Cruz, CA 95064.}

\received{---------------}
\accepted{---------------}

\begin{abstract}
We assess models for the assembly of supermassive black holes (SMBHs)
at the center of galaxies that trace their hierarchical build-up far up in the
dark halo `merger tree'. Motivated by the recent discovery of luminous quasars 
around redshift $z\approx 6$ -- suggesting a very early assembly epoch -- and by 
numerical simulations of the fragmentation of primordial
molecular clouds in cold dark matter cosmogonies, we assume that the first
`seed' black holes (BHs) had intermediate masses
and formed in (mini)halos collapsing at $z\sim 20$ from high-$\sigma$ density 
fluctuations. As these pregalactic 
holes become incorporated through a series of mergers into larger and larger 
halos, they sink to the center owing to dynamical friction, accrete a fraction of the
gas in the merger remnant to become supermassive, form a binary system, and eventually
coalesce. The merger history of dark matter halos and
associated BHs is followed by cosmological Monte Carlo realizations
of the merger hierarchy from early times until the present in a $\Lambda$CDM 
cosmology. A simple model, where quasar activity is driven by major mergers 
and SMBHs 
accrete at the Eddington rate a  mass that scales with the fifth power of the 
circular velocity of the host halo, is shown to reproduce the observed
luminosity function of optically-selected quasars in the redshift range 
$1<z<5$.    
A scheme for describing the hardening of a BH binary in a 
stellar background with core formation due to mass ejection is applied, where
the stellar cusp $\propto r^{-2}$ is promptly regenerated after every major
merger event, replenishing the mass displaced by the binary.
Triple BH interactions will inevitably take place at early times if the 
formation route for the assembly of SMBHs goes back to the very first generation 
of stars, and we follow them in our merger tree. The assumptions underlying 
our scenario lead to the prediction of a population of massive BHs wandering in 
galaxy halos and the intergalactic medium at the present epoch, and 
contributing $\lta 10\%$ to the total BH mass density, $\rho_{\rm 
SMBH}=4\times 10^5\,\mden$ ($h=0.7$). The fraction of binary SMBHs 
in galaxy nuclei is of order 10\% today, and it increases with redshift
so that almost all massive nuclear BHs at early epochs are in binary systems.
The fraction of binary quasars (both members brighter than 0.1 $L_*$) instead
is less than 0.3\% at all epochs. The nuclear SMBH occupation fraction 
is unity (0.6) at the present epoch if the first seed BHs were 
as numerous as the 3.5-$\sigma$ (4-$\sigma$) density peaks at $z=20$.
\end{abstract}

\section{Introduction}
Dynamical evidence indicates that supermassive black holes (SMBHs) reside at the center of most nearby
galaxies (Richstone \etal 1998). The available data show an empirical
correlation between bulge luminosity and black hole mass
(Magorrian \etal 1998), which becomes remarkably tight when the stellar velocity
dispersion of the host bulge, $\sigma_c$, is plotted instead of luminosity
(Ferrarese \& Merritt 2000; Gebhardt \etal 2000). The $m_{\rm BH}$-$\sigma_c$
relation implies a rough proportionality between SMBH mass and the mass of the
baryonic component of the bulge. It is not yet
understood if this relation was set in primordial structures, and consequently how
it is maintained throughout cosmic time with such a small dispersion, or indeed
which physical processes established such a correlation in the
first place (e.g., Silk \& Rees 1998; Haehnelt \& Kauffmann 2000; Adams, Graff,
\& Richstone 2001; Burkert \& Silk 2001). Most recently, it has been shown by
Ferrarese (2002) that in elliptical and spiral galaxies the bulge velocity dispersion
correlates tightly with the value of the circular velocity measured well beyond the
optical radius, suggesting that $m_{\rm BH}$ is actually determined by the mass of
the host dark matter halo.

The strong link between the masses of SMBHs and the gravitational potential wells
that host them suggests a fundamental mechanism for assembling black holes and
forming spheroids in galaxy halos. In popular cold dark matter (CDM)
`bottom-up' cosmogonies, small-mass subgalactic systems form first to merge
later into larger and larger structures. Galaxy halos then experience multiple
mergers during their lifetime, with those between comparable-mass systems
(``major mergers'') expected to result in the formation of elliptical galaxies
(see, e.g., Barnes 1988; Hernquist 1992). Simple models in which SMBHs are also
assumed to grow during major mergers and to be present in every galaxy at any
redshift -- while only a fraction of them is `active' at any given time -- have
been shown to explain many aspects of the observed evolution of quasars (e.g.
Cattaneo, Haehnelt, \& Rees 1999; Cavaliere \& Vittorini 2000; Kauffmann \& Haehnelt
2000; Wyithe \& Loeb 2002). In hierarchical structure formation scenarios, the ubiquity 
of SMBHs in nearby luminous galaxies can arise even if only a small fraction of halos harbor
SMBHs at high redshift (Menou, Haiman, \& Narayanan 2001).
Yet several important questions remain unanswered, most notably: 

1. Did the first 
massive black holes (BHs) form in subgalactic units far up in the merger 
hierarchy, well before the bulk of the stars observed today? 
The seeds of the recently discovery $z\approx 6$ quasars in the Sloan 
Digital Sky Survey (SDSS, Fan \etal 2001b) had to appear at very high redshift,   
$z\gta 10$, if the SDSS quasars are accreting no faster than the Eddington rate and
are not gravitationally lensed or beamed (Haiman \& Loeb 2001).   

2. How massive were the initial BH seeds? A clue to this question might lie in the
numerous population of ultraluminous off-nuclear (`non-AGN') X-ray sources that have 
been detected in nearby galaxies (e.g. Colbert \& Mushotzky 1999; Makishima et al. 2000;
Kaaret et al. 2001).  Assuming isotropic emission, the inferred masses of these `ULXs' 
often suggest intermediate-mass BHs with $m_\bullet\gta$ a few hundred $\msun$.

3. How efficiently do 
SMBHs and their `seeds' spiral inwards and coalesce as they get incorporated
through a series of mergers into larger systems? And what is the cumulative dynamical
effect of multiple BH mergers on galaxy cores? 

4. Is there a population of
relic `Population III' massive holes lurking in present-day galaxy halos?  

In this paper we explore a formation route for the assembly of SMBHs in the nuclei of
galaxies that traces their seeds back to the very first generation of stars,
in (mini)halos above the cosmological Jeans mass collapsing at $z\sim 20$ from the
high-$\sigma$ peaks of the primordial density field. The first stars 
must have formed out of metal-free gas, with the lack of an efficient cooling 
mechanism possibly leading to a very top-heavy initial stellar mass function (IMF;
Larson 1998), and in particular to the production of `very massive stars' (VMSs)
with $m_\star>100\,\msun$ (Carr, Bond, \& Arnett 1984). Recent numerical
simulations of the fragmentation of primordial clouds in standard CDM theories
all show the formation of Jeans unstable clumps with masses exceeding a few hundred
solar masses; because of the slow subsonic contraction -- a regime set up by the
main gas coolant, molecular hydrogen -- further fragmentation into
sub-components is not seen (Bromm, Coppi, \& Larson 1999, 2002; Abel, Bryan, \& 
Norman 2000). Moreover, the different conditions of temperature and density of the 
collapsing cloud result in a mass accretion rate
over the hydrostatic protostellar core $\sim 10^3$ times larger than what observed in 
local 
forming stars, suggesting that Pop III stars were indeed very massive (Omukai \& Nishi 
1998; Ripamonti et al. 2002). If VMSs form above
260 $\msun$, after 2 Myr they would collapse to massive BHs
containing at least half of the initial stellar mass (Fryer, Woosley, \& Heger
2001), i.e., with masses intermediate between those of the stellar and supermassive
variety. It has been suggested by Madau \& Rees (2001, hereafter MR) that a numerous
population of massive BHs may have been the endproduct of the first episode of 
pregalactic star formation; since they form in high-$\sigma$ rare
density peaks, relic massive BHs with $m_\bullet\gta 150\,\msun$
would be predicted to cluster in the cores of more massive halos formed by 
subsequent mergers.

In this paper we expand upon the original suggestion of MR and assess a 
model for the assembly of SMBHs in the nuclei of luminous galaxies out of
accreting Pop III seed holes. The merger history of dark matter halos 
and associated black holes is followed through Monte Carlo realizations of the merger 
hierarchy (merger trees). Merger trees are a powerful tool for tracking  
the evolution of SMBH binaries along cosmic time, and analyze how their 
fate is influenced by the environment (e.g stellar density cusps). We study the 
conditions under which pregalactic massive holes 
may sink to the halo center owing to dynamical friction, accrete a fraction of
the gas in the merger remnant to become supermassive, form a binary system, and
eventually coalesce. Major mergers are frequent at early times, so a 
significant number of binary SMBH systems is expected to form then.
To anticipate the results of our analysis, we find that a simple model where quasar 
activity is driven by major mergers and SMBHs accrete at the Eddington rate a  
mass that scales with the fifth power of the circular velocity of the host halo, 
is able to reproduce the observed luminosity function of optically-selected quasars 
in the redshift range $1<z<5$. Minor mergers are largely responsible for a
population of isolated BHs wandering in galaxy halos, while intergalactic BHs
will be produced by the gravitational slingshot -- the ejection of one BH when 
three holes interact.

\section{Halo merger tree}

There are now a number of algorithms for constructing merger trees, 
the difficulties and drawbacks of various techniques having been 
reviewed by Somerville \& Kolatt (1999). We have developed a Monte Carlo 
algorithm similar in spirit to the one described 
by Cole \etal (2000), and based on the extended Press-Schechter formalism (EPS). 
This gives the fraction of mass in a halo of mass $M_0$ at redshift $z_0$,
which at an earlier time was in smaller progenitors of mass in the range $M$ 
to $M+dM$,
\beq
f(M,M_0)dM={1\over \sqrt{2\pi}}\,{D\over S^{3/2}}\,\exp\left[-{D^2\over 2S}\right]\,
{d\sigma_M^2\over dM}\,dM \label{dn/dm} 
\eeq
where $D\equiv \delta_c(z)-\delta_c(z_0)$ and $S\equiv \sigma_M^2(z)-
\sigma_{M_0}^2(z_0)$ (Bower 1991; Lacey \& Cole 1993). Here $\sigma^2_M(z)$ 
and $\sigma_{M_0}^2(z_0)$ are the linear theory rms density fluctuations smoothed 
with a `top-hat' filter of mass $M$ and $M_0$ at redshifts $z$ and $z_0$,
respectively. The $\delta_c(z)$ and $\delta_c(z_0)$ are the critical thresholds on 
the linear overdensity for spherical collapse at the two redshifts.
Integrating this function over the range $0<M<M_0$ gives unity: all the mass 
of $M_0$ was in smaller subclumps at an earlier epoch $z>z_0$.  
Taking the limit $z\rightarrow z_0$ and multiplying by the factor $M_0/M$ to convert
from mass weighting to number weighting, equation 
(\ref{dn/dm}) gives the number of progenitors the more massive halo fragments
into when one takes a small step $\delta z$ back in time,
\beq
{dN\over dM}(z=z_0)={1\over \sqrt{2\pi}}~{M_0\over M}\,{1\over S^{3/2}}\,
{d\delta_c\over dz}\,{d\sigma_M^2\over dM} \delta z \label{dn/dmdt}. 
\eeq
Our algorithm uses this expression to build a binary merger tree that starts from the
present day and runs backward in time `disintegrating' a parent halo into its 
progenitors. Because for CDM-like power spectra the number of halos
diverges as the mass goes to zero, it is necessary to introduce a cut-off mass
or effective mass resolution $M_{\rm res}$, which marks the transition from 
\emph{progenitor} -- all halos with $M>M_{\rm res}$ -- to 
\emph{accreted mass} -- the cumulative contribution of all halos with $M<M_{\rm res}$
(Somerville \& Kolatt 1999). Having specified the mass resolution, one can 
compute the mean number of fragments in the range $M_{\rm res}<M<M_0/2$,
\begin{equation}
N_p =\int_{M_{\rm res}}^{M_0/2} \frac{dN}{dM}\,dM,
\label{NP}
\end{equation}
and the fraction of accreted mass,
\beq
F_a = \int_{0}^{M_{\rm res}}\frac{dN}{dM}\,{M\over M_0}dM.
\label{eq:fa}
\eeq
Both of these quantities are proportional to the timestep $\delta z$, which is
chosen to ensure that multiple fragmentation is unlikely, i.e., $N_p\ll 1$. 
Hence it is the frequency of mergers that directly controls the 
timestep: this
enables the algorithm to follow the merger process with high time resolution. 
Following Cole \etal (2000), at every timestep a random number $0\leq R \leq 
1$ is generated and compared to $N_p$. If $R \geq N_p$ the parent halo does 
not fragment in this timestep, but its mass is reduced to account for the 
accreted matter, i.e., a new halo is produced with mass $M_0(1-F_a)$. 
Fragmentation occurs instead if $R<N_p$: then
a random value of $M$ in the range $M_{\rm res}<M<M_0/2$ is generated 
from the distribution in equation (\ref{dn/dmdt}), to produce two new halos 
with masses $M$ and $M_0(1-F_a)-M$. The merger hierarchy is built up by 
repeating the same procedure on each subclump at successive timesteps.

To fully define the tree we need to specify the power-spectrum of density fluctuations,
which gives the function $\sigma_M$, and the cosmological parameters $\Omega_0$ and 
$\Omega_\Lambda$, which the critical overdensity for collapse 
$\delta_c$ depends upon.
Unless otherwise stated, all the results shown below refer to the currently favoured
(by a variety of observations) $\Lambda$CDM world model with $\Omega_0=0.3$, 
$\Omega_\Lambda=0.7$, $h=0.7$, $\Omega_bh^2=0.02$, and $n=1$. 
In this cosmology the redshift dependence of the matter density parameter is
$\Omega(z)=\Omega_0(1+z)^3[1-\Omega_0+(1+z)^3\Omega_0]^{-1}$, and the linear
theory growth factor is accurately approximated by       
\beq
D(z)={5\Omega\over 2(1+z)}\left[{1\over 70}+{209\over 140}\Omega-{\Omega^2\over
140}+\Omega^{4/7}\right]^{-1}
\eeq
(Carroll, Press, \& Turner 1992), so that $\sigma_M(z)=\sigma_M(0)D(z)/D(0)$. 
The normalization of the mass fluctuation spectrum, derived from the abundance 
of X-ray emitting clusters observed in the local universe, is 
$\sigma_8(0)\equiv\sigma_0(r=8\,h^{-1}\,\rm Mpc)=0.93$ (Eke, Cole, \& Frenk 1996). 
We have used the fit to the CDM power-spectrum given by Bardeen \etal (1986), 
modified to account for the effects of baryon density following Sugiyama (1995). 
For the spherical collapse density threshold $\delta_c$ we use the fit (accurate
within 0.1\% for $1-\Omega>0.01$) from Nakamura \& Suto (1997),
$\delta_c(z)=1.686[1+0.012299\log (1-\Omega)]$. 

All that remains to be fixed is the mass resolution and timestep.
We have taken $M_{\rm res}$ to represent at $z=0$ a subclump with virial 
velocity 
equal to 10\% that of the $M_0$ parent halo, i.e., $M_{\rm res}=10^{-3}M_0$: 
the mass resolution is taken to decrease with redshift as $(1+z)^{-3.5}$, so 
it is always less than 5\% 
of the mass of the main halo in the merger hierarchy. This ensures a sufficiently 
wide range of masses to allow both for minor and major mergers 
in the tree at all redshifts. Due to the binary nature of the 
algorithm (which allows only a single fragmentation within a timestep), $M_{\rm res}$ 
cannot be chosen to be arbitrarily small, as in this case
multiple fragmentation becomes unprobable (i.e., $N_p\ll 1$) only if the time 
resolution (and hence the computational time) is extremely high. 
For instance, if we keep the ratio between $M_{\rm res}$ and the mass of the main 
halo in the tree fixed at the present-day value, the time for one Monte Carlo
realization is longer by an order of magnitude.
Since our aim is to track the merger hierarchy to very high redshifts, to keep
the computational time down to acceptable values, we have approximated the 
fraction of accreted mass by the fitting formula
\beq
F_a=a[\log(M/M_{\rm res})]^{-b},
\eeq
instead of integrating numerically the mass function of progenitors with 
$M<M_{\rm res}$ at every branch of the tree. Here, depending on 
mass and redshift, $a$ and $b$ are in the range $(0.3-40)\times10^{-3}$ 
and $0.3-0.8$, 
respectively, and have been adjusted empirically to provide a fit to 
equation 
(\ref{eq:fa}) within 5\% in the range of masses and reshifts considered.
Beside conserving mass, a merger tree algorithm must reproduce at all redshifts the 
conditional mass function predicted by the EPS theory. The 
comparison is shown in Figures \ref{fig1} and \ref{fig2}: for this set of 
realizations we have used 820 timesteps logarithmically spaced in expansion factor 
between $z=0$ and $z=20$. With the above prescriptions, the tree typically agrees 
with the EPS predictions within a factor of 2 up to $z=20$ for masses greater than 
3$\times M_{\rm res}$. Each realization we generate tracks backwards the merger 
history of 220 parent halos at the present epoch picked over the mass 
range $10^{11}<M_0<10^{15}\,\msun$, where the lower limit has been chosen to 
match the minimum mass effectively probed by dynamical studies of SMBH hosts in 
the local universe.

\begin{figurehere}
\vspace{+0.5cm}
\centerline{
\psfig{file=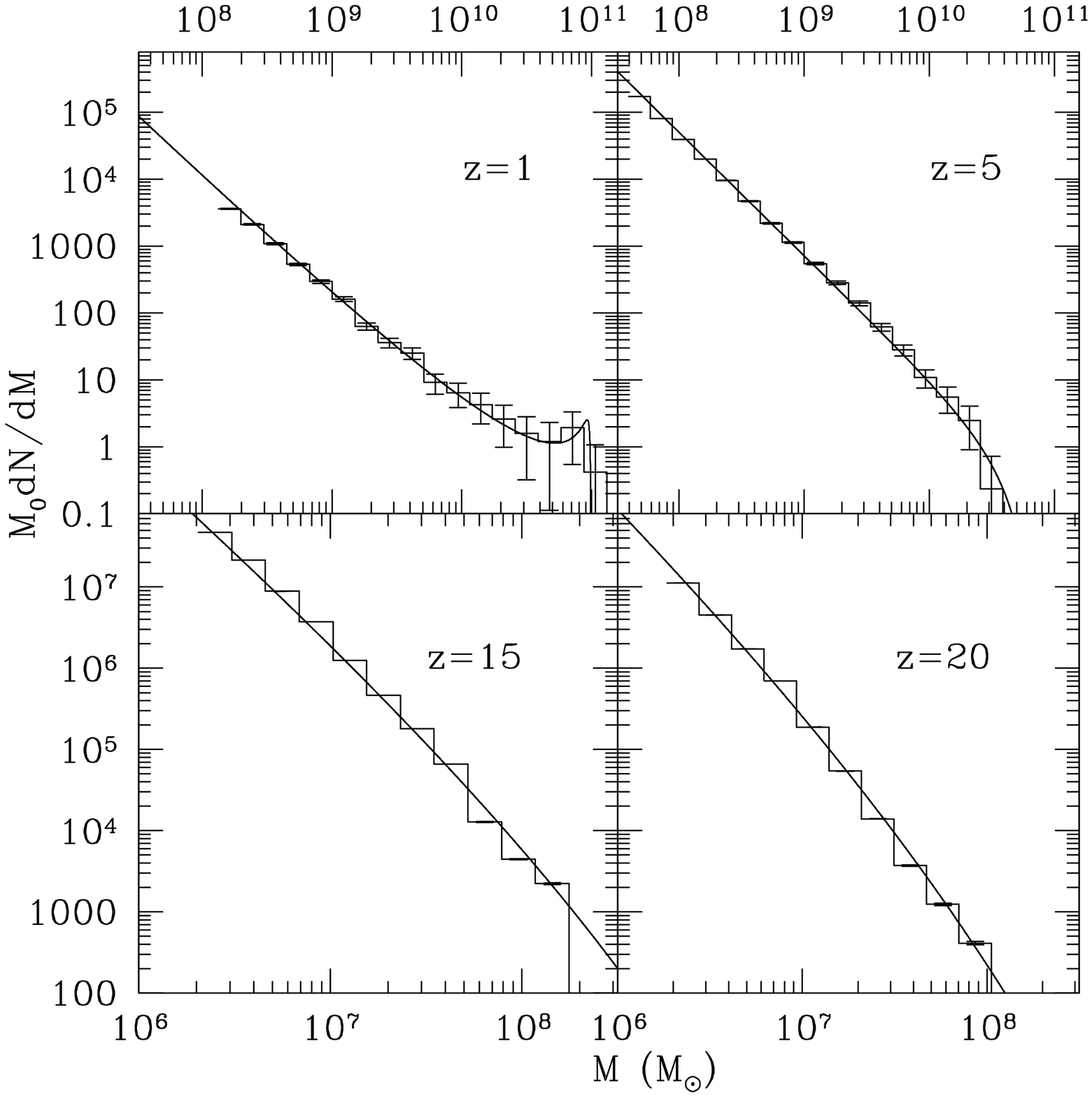,width=2.9in}}
\caption{\footnotesize Mean number of progenitors with mass $M$ for a $z_0=0$, 
$M_0=10^{11}\,\msun$ parent halo, at redshifts $z=1,5,15,20$. 
{\it Solid lines:} predictions of the EPS theory. {\it Histograms:} results 
for the merger tree (mean of 50 realizations), $M>2\times M_{\rm res} $ . Error 
bars represent the Poissonian error in the counts.}
\label{fig1}
\end{figurehere}
\vspace{+0.4cm}

These are broken up into as many as 70,000 progenitors by $z=20$. 
One issue of concerns involves systematic
deviations of the unconditional (and conditional) Press-Schechter (PS) mass function compared 
with N-body simulations. At low redshifts, the PS mass function overpredicts the 
number of small halos by a factor of 1.5 to 2 (e.g. Gross \etal 1998; Sheth \& 
Tormen 1999). The model and simulation results agree well on all scales at $z\sim 1$,
while at higher redshifts the abundance of large-mass halos is underestimated by the
PS model (Somerville \etal 2000). In general, halo merging histories constructed
using the EPS formalism are reasonably consistent with those extracted from 
N-body simulations. At very high redshift ($z\gta 10$), a recent cosmological 
simulation by Jang-Condell \& Hernquist (2001) finds good agreement with the PS
mass function for the lowest halo masses of interest here. 

\begin{figurehere}
\vspace{+0.5cm}
\centerline{
\psfig{file=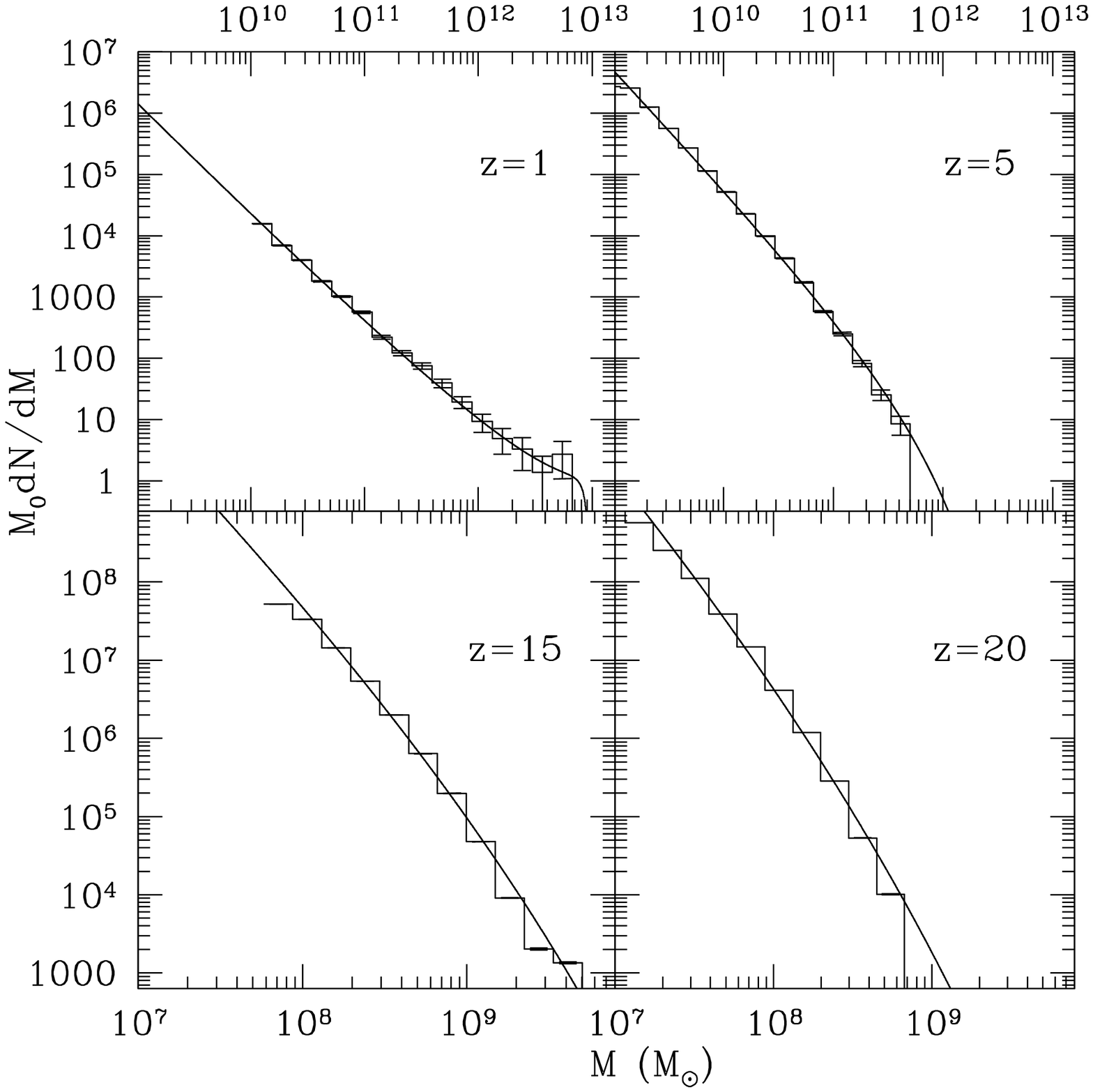,width=2.9in}}
\caption{\footnotesize Same as Fig. 1 for a $z_0=0$, $M_0=4\times 10^{13}\,\msun$ 
parent halo.}
\label{fig2}
\end{figurehere}

\section{Growth of supermassive black holes}

\subsection{Pregalactic `seeds'}

Following MR, we will assume that one seed BH forms in each of the rare density
peaks above $\nu$-$\sigma$ at $z=20$. All seed BHs are assigned a mass 
$m_\bullet=150\,\msun$. In general, we find that our results are not very sensitive 
to the precise mass choice for these seed holes (see \S\,4). As our fiducial model 
we take $\nu=3.5$, corresponding 
in the assumed $\Lambda$CDM cosmology to minihalos of mass $M_{\rm seed}=1.1\times 
10^7\,h^{-1}\,\msun$. This is larger than the minimum mass threshold for baryonic 
condensation, $M_{\rm min}\approx 5\times 10^5\,\msun$ ($h=0.65$), found in the 
numerical simulations of Fuller \& Couchman (2000). Above $M_{\rm min}$ the H$_2$ 
cooling time is shorter than the Hubble time at virialization, the gas in the 
central halo regions becomes self-gravitating, and stars can form. 
We notice that the EPS formalism predicts of order 1 progenitor above $M_{\rm seed}$ 
for our lower mass parent halo, $M_0=10^{11}\,\msun$. Also, our resolution mass,
$M_{\rm res}=10^{-3}M_0(1+z)^{-3.5}$ is always lower at $z=20$ than $M_{\rm seed}$
for all parent halos with masses $M_0<10^{15}\,\msun$. 

A pregalactic halo at $z=20$ is characterized by a virial radius (defined as the radius of the 
sphere encompassing a mean mass density $\Delta_\vir\,\rho_{\rm crit}$, where 
$\rho_{\rm crit}$ is the critical density for closure at the redshift $z$ and
$\Delta_\vir$ is the density contrast at virialization\footnote{For the assumed 
cosmology this can be approximated by $\Delta_\vir=178\Omega^{0.45}$ (Eke, 
Navarro, \& Frenk 1998).}) 
$r_\vir=390\,{\rm pc}~M_7^{1/3}\,h^{-1}$, and a circular velocity $V_c=10.5\,
{\kms}~M_7^{1/3}$ at $r_\vir$. 
The gas collapsing along with the dark matter perturbation will be 
shock heated to the virial temperature $T_\vir\approx 3200\,{\rm K}~M_7^{2/3}\,$,
where $M_7$ is the halo mass in units of $10^7 h^{-1}\,\msun$. The total 
baryonic mass within the virial radius is equal to $(M_{\rm seed}\Omega_b/\Omega_0)$.
In a Gaussian theory, halos more massive than the $\nu$-$\sigma$ peaks contain 
a fraction erfc($\nu/\sqrt{2}$) ($=0.00047$ for $\nu=3.5$) of the mass of the 
universe. Therefore the mass density parameter of our `3.5-$\sigma$' pregalactic 
holes is
\beq
\Omega_\bullet={0.00047\,\Omega_0\,m_\bullet\over 1.1\times 10^7\,h^{-1}\,
\msun}\gta 2\times 10^{-9}\,h.
\eeq
This is much smaller than the density parameter
of the supermassive variety found in the nuclei of most nearby galaxies,
\beq
\Omega_{\rm SMBH}={6\times 10^5\,h\, \mden\over 2.8\times 10^{11}\,h^2\mden}
\approx 2\times 10^{-6}\,h^{-1},
\eeq
where the value at the denominator is the critical density today and we have taken
for the local density of SMBHs the value recently inferred by Merritt \& Ferrarese
(2001). Clearly, if SMBHs form out of very rare Pop III BHs, the present-day mass 
density of SMBHs must have been accumulated 
during cosmic history via gas accretion, with BH-BH mergers playing a secondary
role. This is increasingly less true, of course, if the seed holes are more numerous 
and populate the 2- or 3-$\sigma$ peaks instead, or halos with smaller masses 
at $z>20$ (MR). 

The choice of where to initially locate our seed BHs, while motivated by 
recent numerical simulations of the formation of the first stars, is 
clearly somewhat arbitrary. Computational costs do not allow us to follow
the merger hierarchy much beyond $z=20$, or down to minihalo masses 
smaller than $M_{\rm seed}$. As argued by MR, if an extreme IMF
is linked to primordial H$_2$ chemistry and cooling, it seems unlikely that
the formation of massive BHs from zero-metallicity VMSs might have been a very
efficient process, since metal-free VMSs are copious sources of Lyman-Werner photons
(Bromm, Kudritzki, \& Loeb 2001). This radiation may photodissociate H$_2$ elsewhere 
within the host halo, escape into the
intergalactic medium, and form a soft UV background that suppresses 
molecular cooling throughout the universe (Haiman, Abel, \& Rees 2000) and 
may inhibit the formation of a much more numerous population of VMSs.  
Moreover, VMSs in the mass range $140\le m_\star\le 260\,\msun$ are predicted
to make pair-instability supernovae with explosion (kinetic) energies of up 
to $10^{53}\,$erg. This is typically
much larger than the baryon binding energy of a subgalactic fragment,
$E_b\approx 10^{52.5}\,{\rm erg}\,h^{-1}\,\Omega_b\,M_{7}^{5/3}\,[(1+z)/20]$.
Minihalos will then be completely disrupted (`blown-away') by such energetic
events, metal-enriched material will be returned to the IGM (e.g. Madau,
Ferrara, \& Rees 2001) and later collect in the cores
of more massive halos formed by subsequent mergers, where a `second' generation
of stars may now be able to form with an IMF that is less biased towards very
high stellar masses (Schneider \etal 2002).

\subsection{Major mergers}
We model each dark halo as a singular isothermal sphere (SIS) with 
circular velocity $V_c$, one-dimensional velocity dispersion 
$\sigma_{\rm DM}=V_c/\sqrt{2}$,
and density $\rho(r)=V_c^2/4 \pi G r^2$, truncated 
at the virial radius. When two halos of mass $M$ and $M_s$ merge, the 
`satellite' (less massive) progenitor (mass $M_s$) is assumed to sink to the 
center of the more massive pre-existing system on the Chandrasekhar dynamical 
friction (against the dark matter background) timescale
\beq
t_{\rm df}=1.17{r_{\rm circ}^2 V_c\over GM_s\ln\Lambda}\epsilon^\alpha
=1.65{1+P\over P}\,{1\over H\sqrt{\Delta_\vir}\ln\Lambda}\Theta
\label{eqtdf}
\eeq
(Lacey \& Cole 1993; Binney \& Tremaine 1987), where $V_c$ is the circular 
velocity of the satellite in the new halo of mass $M+M_s$ and virial radius
$r_\vir$, $r_{\rm circ}$ is the radius of 
the circular orbit having the same energy as the actual orbit, the `circularity' 
$\epsilon$ is the ratio between the orbital angular momentum and that of the 
circular orbit having the same energy, $H$ is the Hubble parameter, $P=M_s/M$ 
is the (total) mass ratio of the progenitors, 
and the Coulomb logarithm is taken to be $\ln\Lambda\approx\ln(1+P)$.
The dependence of this timescale on the orbital parameters is contained in the
term
\beq
\Theta=\epsilon^\alpha (r_{\rm circ}/r_\vir)^2.
\label{orbit}
\eeq
The most likely orbits occurring in cosmological CDM simulations of structure 
formation have circularity $\epsilon=0.5$ and $r_{\rm circ}/r_\vir=0.6$ (e.g. 
Tormen 1997; Ghigna \etal 1998). With these initial orbital parameters,
recent numerical investigations by van den Bosch \etal
(1999) and  Colpi, Mayer, \& Governato (1999) suggest a value $\alpha=0.4$-0.5 
for the exponent in equation (\ref{orbit}). Here we assume $\Theta=0.3$, but we note 
that the merger timescale computed in this way does not include the increase in the
orbital decay timescale due to tidal stripping of the satellite (Colpi \etal 1999).  
Satellites will merge with the central galaxy on timescales shorter than the
then Hubble time only in the case of major mergers, $P\gta 0.3$. In minor 
mergers tidal stripping may leave the satellite BH almost `naked' of its dark halo,
too far from the center of the remnant for the formation of a black hole binary.

Figure \ref{fig3} shows the number of major
mergers per unit redshift bin experienced by halos of different masses. For 
galaxy-sized halo this quantity happens to peak in the redshift range 2-4, the 
epoch when the observed space density of optically-selected quasar also reaches a 
maximum. Hydrodynamic simulations of major mergers have 
shown that a significant fraction of the gas in interacting galaxies falls to 
the center of the merged system (Mihos \& Hernquist 1994, 1996): the cold gas may be
eventually driven into the very inner regions, fueling an accretion episode
and the growth of the nuclear BH. In the following we shall make the simplifying 
assumption that SMBHs accrete material only during major mergers.

\begin{figurehere}
\vspace{+0.5cm}
\centerline{
\psfig{file=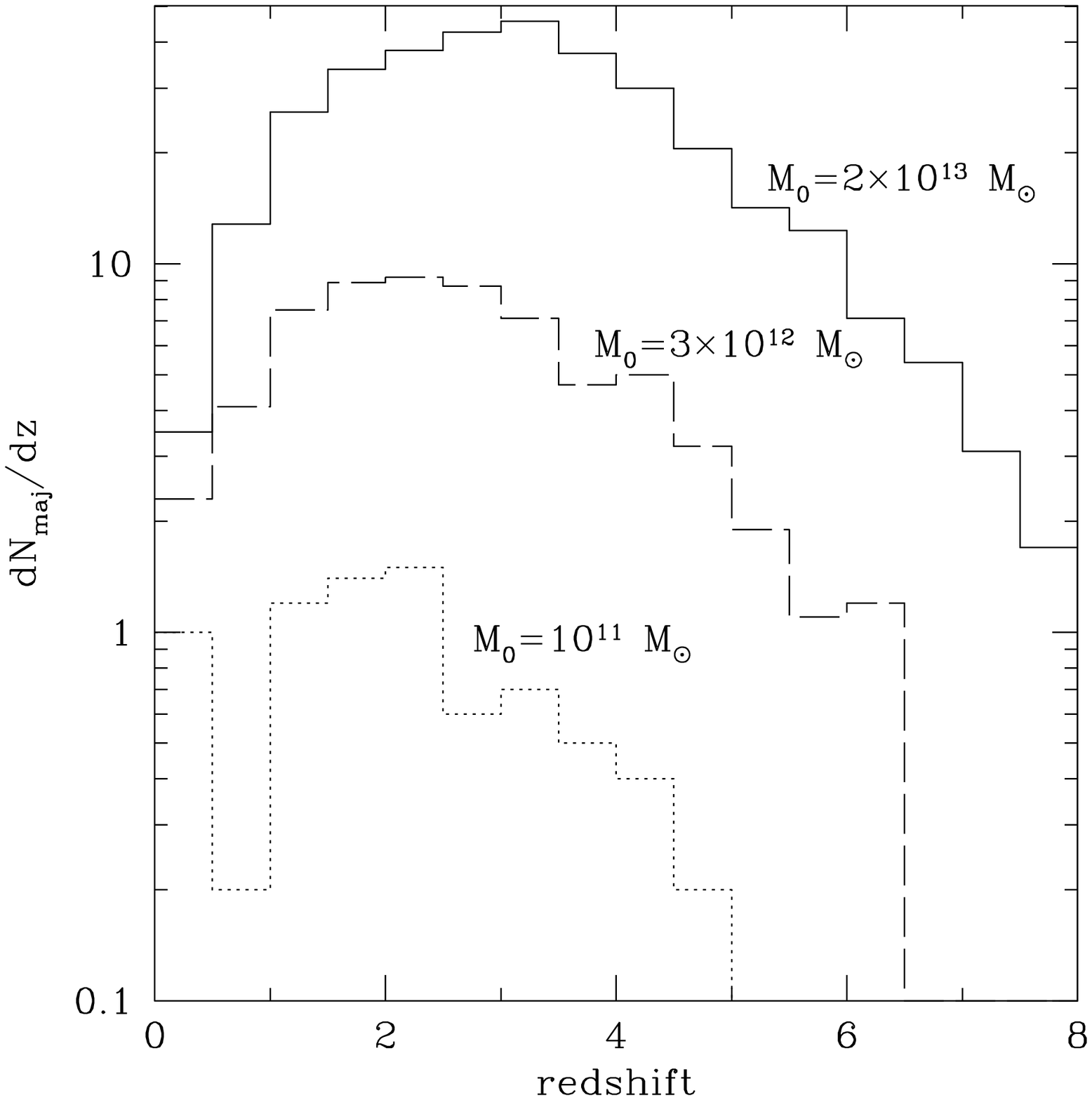,width=2.6in}}
\caption{\footnotesize Mean number of major mergers experienced per unit 
redshift by halos with masses $>10^{10}\,\msun$. {\it Solid line:} 
progenitors of a $M_0=2\times 10^{13}\,\msun$ halo at $z=0$. {\it Dashed line:}
same for $M_0=3\times 10^{12}\,\msun$. {\it Dotted line:} same for
$M_0=10^{11}\,\msun$.}
\label{fig3}
\end{figurehere}

\subsection{Accretion history}
The physical processes that determine the amount of accreted gas and the
characteristic accretion timescales onto SMBHs are poorly understood, and different 
prescriptions have been proposed in the literature to explain the 
observed evolution of QSOs within hierarchical clustering cosmologies.
We shall not attempt here to model these processes in details, but we note that the
fraction of cold gas ending up in the hole must depend on the properties of the host
halo in such a way to ultimately lead to the observed correlation between stellar 
velocity dispersion and SMBH mass. Using 
the most up-to-date set of black hole mass measurements, Ferrarese (2002) finds
\beq
m_{\rm BH}=(4.4\pm 0.9)\times 10^7\,\msun~\sigma_{c,150}^{4.58\pm 0.52},
\label{msigma}
\eeq
where $\sigma_{c,150}$ is the bulge velocity dispersion (defined within an 
aperture of size $\lta 0.5\,$kpc) in units of $150\,\kms$. Gebhardt \etal (2000) 
and Tremaine \etal (2002) report a similar relation with a somewhat shallower slope. 
From a sample of 
spirals and elliptical galaxies with $\sigma_c>70\,\kms$, Ferrarese (2002) also 
shows that the stellar velocity dispersion is strictly correlated with the 
asymptotic value of the circular velocity $V_c$ measured well beyond the optical
radius,
\beq
\log V_c=(0.88\pm 0.17)\log \sigma_c+(0.47\pm 0.35). \label{ferr2002}
\eeq
To avoid introducing additional parameters to our model, as well as 
uncertainties linked to gas cooling, star formation, and supernova feedback, 
we combine the two previous relations and adopt the following simple 
prescription for the mass accreted by a SMBH during each major merger:
\begin{equation}
\Delta m_{\rm acc}=1.3\times 10^4\,\msun~{\cal K}\,V_{c,150}^{5.2},
\label{macc_eq}
\end{equation}
where $V_{c,150}$ is the circular velocity of the merged system in units of 
150 $\kms$ (cf. Kauffmann \& Haehnelt 2000). We assume this mass is accreted 
by the BH in the more massive progenitor halo at 
the Eddington rate,  
after about a dynamical timescale (estimated at one tenth the virial radius, 
$t_{\rm dyn}=0.1 r_\vir/V_c$). 
The normalization factor $\cal K$ is of order unity and is fixed 
in order to reproduce the $m_{\rm BH}-\sigma_c$ relation observed locally.

The above relations are only valid up to a value of the velocity dispersion 
corresponding to galaxy group and cluster scales. For instance, the SMBH in M87 has 
a mass of $m_{\rm BH}=3\times 10^{9}\,\msun$ (Harms et al. 1994), perfectly 
correlating with the  observed circular velocity $V_c=506 \kms$ of the galaxy, but
not with the circular velocity of the Virgo cluster ($V_c \approx 1000 \kms$).
Moreover, the larger halo ($V_c=1280 \kms$) considered in our merger-tree set
would contain at $z=0$ a SMBH more massive than $10^{10}\,\msun$: this would lead
to an overestimate of the quasar LF at late epochs. 
Following Kauffmann \& Haehnelt (2000), we have then inhibited gas accretion onto 
SMBHs in all halos with $V_c>600\kms$. We find that this assumption affects only 
the accretion history of the SMBHs hosted in the two more massive halos of our 
realizations, and only at late epochs. On small mass scales, on the other hand,
accretion  onto BHs hosted by minihalos with virial temperature $T_{\rm vir}<10^4\,$
K may be inhibited by UV radiation. Photoionization by an internal
UV source or a nearby external one will heat the gas inside these shallow potential
wells to $10^4\,$K, leading to the photoevaporation of baryons out of their hosts. 
After the reionization epoch gas cooling, star formation, and accretion onto BHs
may only possible within more massive halos with virial temperatures $T_\vir\gta$ a 
few $\times 10^4\,$K, where pressure support is reduced and gas can condense due to 
atomic line processes. In our model we therefore suppress gas accretion onto all BHs 
in minihalos with $T_{\rm vir}<10^4\,$. We have checked that this assumption has 
little or no effect at $z\lesssim 10$. 

Modeling gas accretion onto BHs with the recipes just described we find that, 
along cosmic history, most of the final mass of SMBHs come from gas accretion, 
rather than from BH merging. The final mass $m$ of the SMBH formed after coalescence 
assumes the entropy-area relation for BHs (maximally efficient radiative merging,
e.g. Ciotti \& van Albada 2001): the total entropy $S$ of the system 
remains unchanged,  
$S=m^2/4=S_1+S_2=m_1^2/4+m_2^2/4$ (taking  $G=c=k=h=1$; Hawking \& Ellis 1973).
In contrast with previous work (e.g. Kauffmann \& Haehnelt 2000; 
Menou \etal 2001), we do not make here the simplifying assumption that the two 
pre-existing holes coalesce instantaneously. The evolution of SMBH pairs will be 
discussed in the next section.

\subsection{Dynamical evolution of BH binaries}

In our model the merging -- driven by dynamical friction against the dark matter 
background -- of two halo$+$BH systems with mass ratio $P\gta 0.3$ will drag in a 
satellite BH towards the center of the more massive progenitor; this will
inevitably lead to the formation of a bound SMBH binary in the violently relaxed 
core of the newly merged stellar system. Figure \ref{fig4} shows the typical 
mass ratio of binary BHs together with the mean mass of the larger member of the pair as a function 
of redshift. At late epochs most of the BH pairs have unequal masses and, 
since the growth history of SMBHs does not track that of dark matter halos, 
a major merger between halos does not necessarily result in a BH binary with a large $P$,
as evident from the upper panel of the figure. 

\begin{figurehere}
\vspace{+0.5cm}
\centerline{
\psfig{file=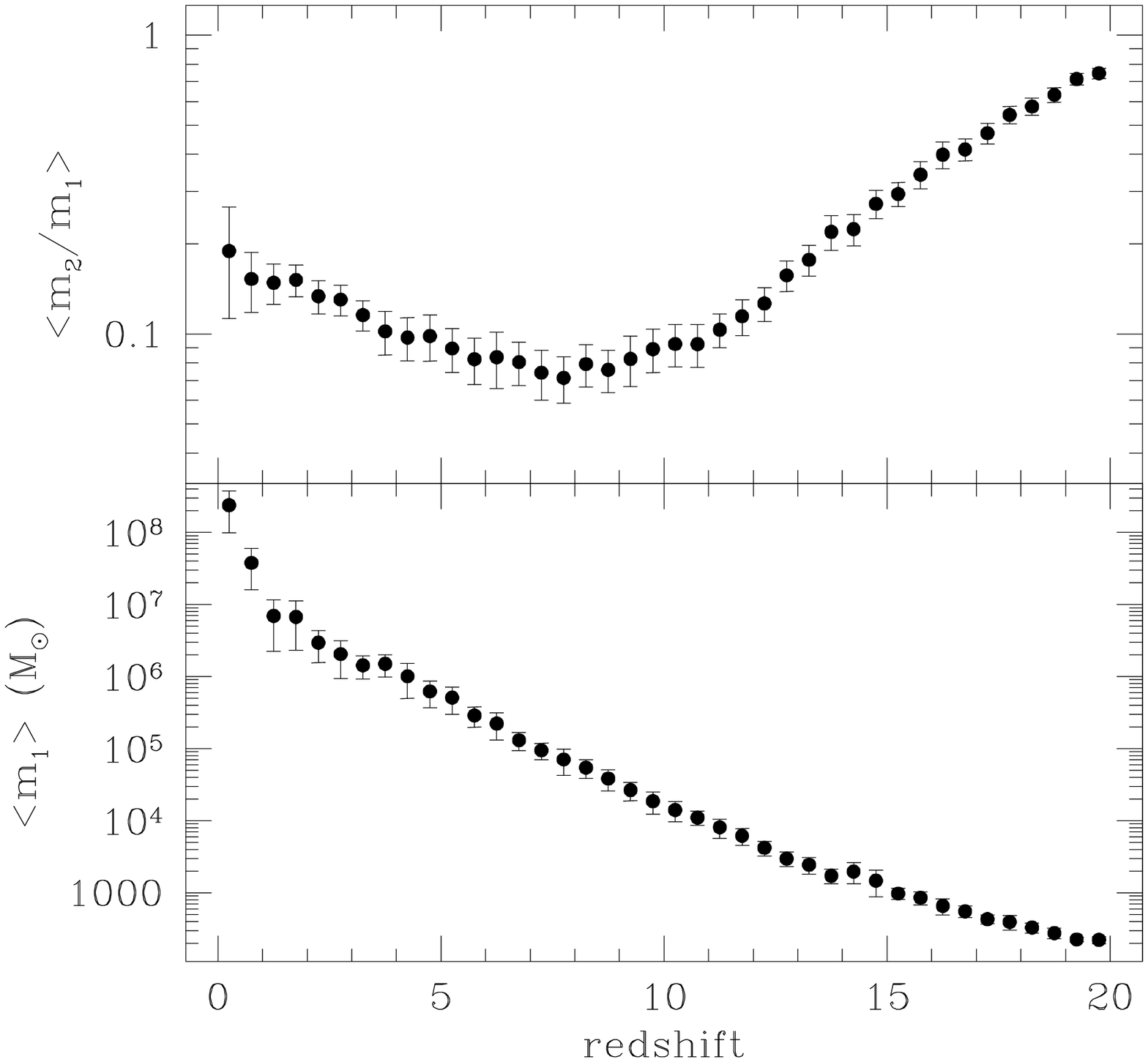,width=2.9in}}
\caption{\footnotesize The mass ratio of BH binaries ({\it upper panel}) 
and the mass of the larger BH ({\it lower panel}) are shown as a function of 
redshift. The points show the mean values and 1-$\sigma$ error bars from all 
Monte Carlo realizations. At high redshift binary members are seed Pop III BHs with 
nearly equal masses; as time goes on the holes grow mainly due to gas accretion, and 
low mass ratios become more probable.}
\label{fig4}
\end{figurehere}
\vspace{+0.4cm}

The subsequent evolution of BH binaries was first outlined by Begelman, Blandford, \& 
Rees (1980). Consider a binary with BH masses $m_1\ge m_2$ and
semimajor axis $a(t)$ in an isotropic background of stars of mass $m_\star\ll m_2$ 
and density $\rho_\star(r)$. We use a simple model for the initial central stellar 
distribution,
an SIS with a velocity dispersion comparable to the halo $\sigma_{\rm DM}$,
\beq
\rho_\star={\sigma_{\rm DM}^2\over 2\pi Gr^2}.
\label{eqSIS}
\eeq
This appears to be a good assumption for early-type lens galaxies (e.g. Koopmans
\& Treu 2002). When the age of the system is larger than the stellar relaxation time, 
the equilibrium distribution of stars around a BH is expected to be cuspy, $\rho_\star\propto 
r^{-7/4}$, within the gravitational sphere of influence of the BH, even if the original 
profile had a core (Bahcall \& Wolf 1976).

The binary 
forms at a separation $a_b=G(m_1+m_2)/(2\sigma_{\rm DM}^2)$ at which the enclosed 
stellar mass equals 
$m_1+m_2$, and initially hardens by dynamical friction from distant stars acting on each 
BH individually. But as the binary separation shrinks (the binary `hardens') the
effectiveness of dynamical friction slowly declines because distant encounters perturb 
only the binary center's of mass but not its semimajor axis. The BH pair then hardens
via three-body interactions, i.e., by capturing the stars that pass 
within a distance $\sim a$ of it and ejecting them (`gravitational slingshot') at 
much higher velocities, 
$v_{\rm ej}\approx V_{\rm bin}\equiv
[G(m_1+m_2)/a]^{1/2}$, where $V_{\rm bin}$ is the relative velocity of the two BHs
if their orbit is circular: this is the hard binary stage. In Quinlan's
(1996) simulations of the dynamical evolution of massive BH binaries, the system 
does not become hard until $a$ falls below
\\
\beq
a_h={Gm_2\over 4\sigma_{\rm DM}^2}=1\,{\rm pc}~~ \left({m_2\over 
10^{7.3}\,\msun}\right)\,\sigma_{\rm DM,150}^{-2}
\eeq
(Quinlan 1996).\footnote{The standard definition of a ``hard'' binary, one where its
binding energy $E_b=Gm_1m_2/a$ exceeds the typical kinetic energy of the surrounding
stars $3m_\star\sigma_{\rm DM}^2/2$ (Binney \& Tremaine 1987), is inapplicable 
to massive BH binaries as they are always hard if bound. Quinlan (1996) defines hardness instead 
in terms of the binary orbital velocity; in his definition a hard binary hardens at a 
constant rate.} We assume that the `bottleneck' stages of the binary shrinking  occur 
for separations $a<a_h$; in a major merger, after a dynamical friction timescale, 
we form  the BH binary at a separation $a_h$ and let it evolve.   

In a fixed background, the hardening timescale $|a/\dot a|$ decreases with $a$,
\begin{equation}
t_h=\frac{\sigma_{\rm DM}} {G \rho_\star a H}=
10^{3.4}\,{\rm yr}\left({a\over {\rm pc}}\right)\left({15\over H}\right) 
\sigma^{-1}_{\rm DM,150}, \label{th}
\end{equation}
and the binary would spend the longest period of time with $a\approx a_h$. Here
the second equality assumes an SIS (eq. \ref{eqSIS}) down to a 
distance $a$ from the center, and
the dimensionless hardening rate is $H\approx 15$ in the limit of a very hard, 
equal-mass binary (Quinlan 1996). If the hardening continues sufficiently far, 
gravitational radiation losses can take over, and the two BHs rapidly coalesce on the 
timescale (for a circular orbit)
\begin{equation}
t_{\rm gr}={5c^5 a^4(t)\over 256 G^3 m_1m_2(m_1+m_2)}
\end{equation}
(Peters 1964). 
If the binary can shrink to a separation
\beq
a_{\rm gr}=0.014\,{\rm pc}~~\left[{(m_1+m_2)m_1m_2\over 10^{21.3}\,\msun^3}\right]^{1/4}, 
\eeq
the binary will coalesce within 10 Gyr due to the emission of gravitational waves. Here 
we have normalized to the case $m_1=m_2=10^7\,\msun$. In our model we neglect the 
recoil due to the non-zero net linear momentum carried away by gravitational waves in 
the coalescence of two unequal mass BHs (`gravitational rocket'). Radiation recoil is a 
strong-field gravitational effect that depends on the lack of symmetry in the system,
and may eject massive BHs from galaxy cores. To date, the outcome of a gravitational
rocket remains uncertain. Newtonian-approximation and perturbative calculations of two
orbiting Schwarzschild holes suggest 
maximum recoil velocities of order $70-100\,\kms$ (Fitchett 1983, Fitchett \& 
Detweiler 1984). 

In practice, however, one cannot assume a fixed stellar background in estimating the 
rate of BH mergers, as the hardening of the binary modifies the stellar density 
$\rho_\star$ in 
equation (\ref{th}):
the shrinking of the pair removes mass interior to the binary orbit, depleting the galaxy 
core of stars and slowing down further hardening. The effect of loss-cone depletion 
(the depletion of low-angular momentum stars that get close enough to extract energy from 
a hard binary) is one of the major uncertainties in computing the merger time, and makes
it difficult to construct viable merger scenarios for BH binaries. 
A recent analysis by Yu (2002) has shown that in significantly flattened or triaxial galaxies
the supply of low-angular momentum stars may be sufficient to reach $a_{\rm gr}$.    
A massive gaseous disk surrounding the binary may further speed up the merger rate (Gould \& Rix
2000). N-body simulations of BH binary decay suggest that the wandering of the binary center of
mass from the 
galaxy center (and to a lesser extent the diffusion of stars into the loss cone) may also work to
mitigate the problems associated with loss-cone depletion (which may ultimately cause the binary
to `stall') and helps the binary merge (Quinlan \& Hernquist 1997; Milosavljevic \& Merritt 2001). 
Here we adopt a simple analytical scheme following
Merritt (2000) that qualitatively reproduces the evolution observed in N-body simulations.
If ${\cal M}_{\rm ej}$ is the stellar mass ejected 
by the BH pair, the binary evolution and its effect on the galaxy core are determined
by the coupled equations
\beq
{d\over dt}\left({1\over a}\right)=H {G\rho_\star\over \sigma_\DM},
\label{eqH}
\eeq
and
\beq
{d{\cal M}_{\rm ej}\over d\ln(1/a)}=J(m_1+m_2),
\label{eqJ}
\eeq
where $J$ is the dimensionless mass ejection rate, $J\approx 1$ nearly independent of
$a$ for $a\ll a_h$ (Quinlan 1996). Integrating the second equation one finds 
${\cal M}_{\rm ej}\approx J(m_1+m_2)\ln(a_h/a)$: the binary ejects of order its own mass 
in shrinking from $a_h$ to $a_h/3$. We assume that the stellar mass removal 
creates a core of radius $r_c$ and constant density $\rho_c\equiv \rho_\star(r_c)$, so that the 
ejected mass can be written as  
\begin{equation}
{\cal M}_{\rm ej}={2\sigma_{\rm DM}^2r_c\over G}-{4\pi\over 3}\rho_c r_c^3= {4\over 3}{\sigma_{\rm DM}^2 
r_c\over G}
\label{rcore}
\end{equation}
(Merritt 2000).  From equations (\ref{eqJ}) and (\ref{rcore}) one derives
\begin{equation}
r_c(t)={3\over4 \sigma_{\rm DM}^2}G J (m_1+m_2)\ln(a_h/a),
\end{equation}
and the core density decreases as
\begin{equation}
\rho_c(t)=\frac{8\sigma_{\rm DM}^6}{9\pi G^3{\cal M}^2_{\rm ej}(t)}.
\label{rc}
\end{equation}
The above relations, assuming a constant $\sigma_{\rm DM}$ during the 
hardening of the binary, are strictly valid only if the stellar relaxation 
timescale is long compared to the hardening time.
The binary separation quickly falls below $r_c$ and subsequent evolution is slowed down
due to the declining stellar density, with a hardening time,
\begin{equation}
t_h=\frac{2\pi r_c^2}{H\sigma_{\rm DM}a},
\label{hardt}
\end{equation}
which now becomes increasingly long as the binary shrinks.  In this model the mass ejected
increases logarithmically with time, and the binary can `heat' background stars at radii $r$,
$a\ll r \lta r_c$.  In N-body simulations this may happen due to the Brownian motion of the binary
induced by continuous interactions with other stars. Also, stars on eccentric orbits are most 
likely to interact with
the binary and be removed, then loose their kinetic energy to the background as they spiral 
back in and are kicked out again.  

Figure \ref{fig5} shows the evolution of a binary of seed, intermediate-mass BHs
in two dark matter halos with different velocity dispersion. The binary 
separation, $a$, shrinks as the pair interacts with the surrounding stellar field,
and, at the same time, 
the ejection of stars decreases the central density creating a stellar core. 

\begin{figurehere}
\vspace{+0.5cm}
\centerline{
\psfig{file=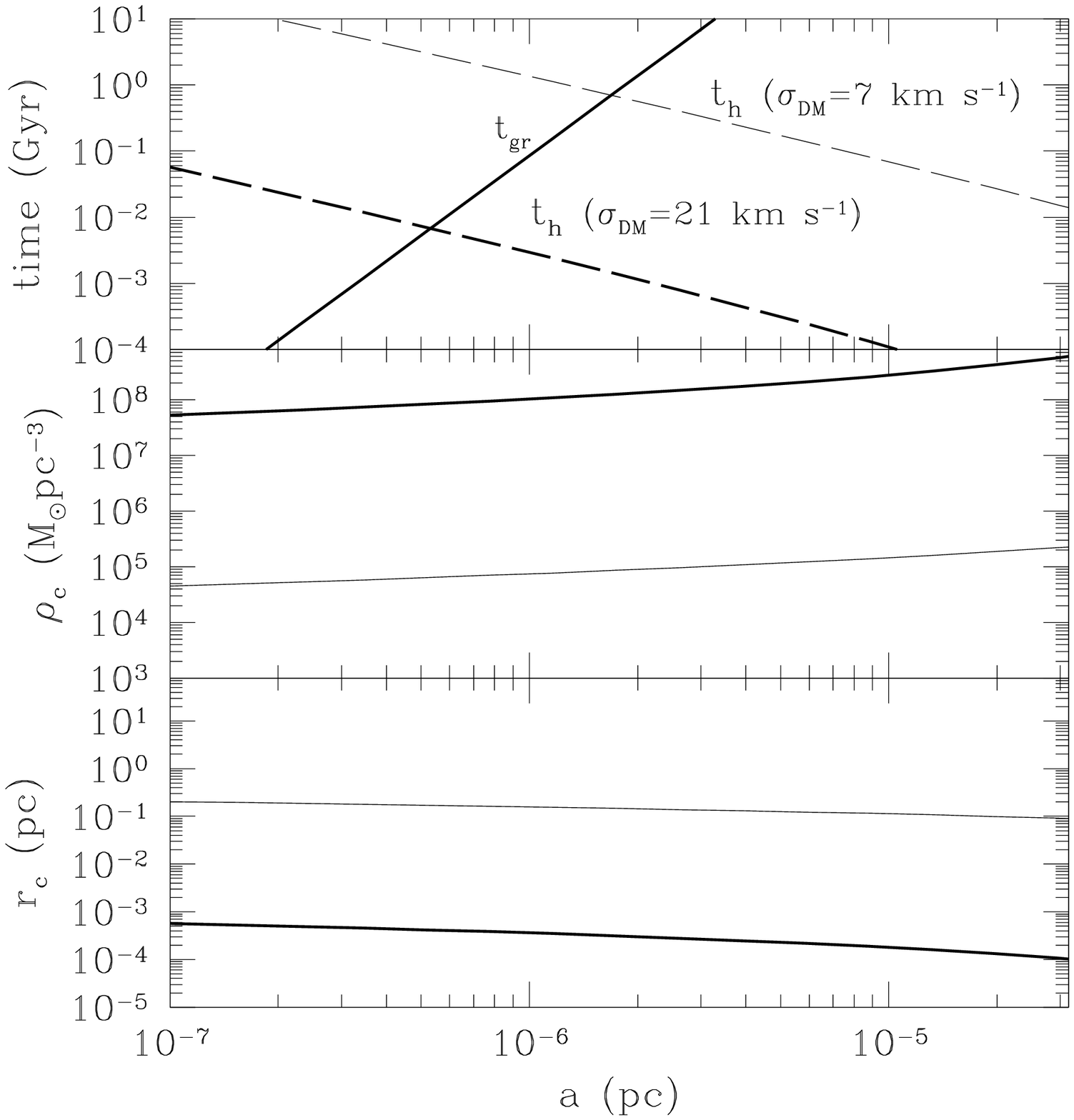,width=2.9in}}
\caption{\footnotesize The evolution (from bottom to top) of core radius, core stellar density,
and hardening timescale during the shrinking of a `Pop III' BH binary against its separation. 
As the pair of $m_\bullet=150\,\msun$ BHs shrinks, the initial $\rho_\star\propto r^{-2}$ stellar 
cusp is gradually converted into a constant density core by the gravitational slingshot, 
and the hardening timescale lengthens. {\it Thin lines:} $\sigma_{\rm DM}=7\,\kms$. 
The total stellar mass ejected prior to coalescence is 7.5$(m_1+m_2)$. 
{\it Thick lines:} $\sigma_\DM=20\,\kms$. The total stellar mass ejected prior to 
coalescence is 6.5$(m_1+m_2)$.} 
\label{fig5}
\end{figurehere}
\vspace{+0.4cm}

Finally, as it is conceivable that major mergers between galaxies may trigger bursts
of star formation (e.g. Somerville, Primack, \& Faber 2001), we further assume that a stellar cusp 
$\propto r^{-2}$ is {\it promptly regenerated after every major merger event}, replenishing the 
mass displaced by the BH binary. 
For a fixed binary mass the coalescence timescale is shorter in the case of more massive 
galaxies.  The evolution of two SMBH binaries in a $\sigma_\DM=200\,\kms$ halo is
depicted in Figure \ref{fig6}.

An equal mass binary with $m_1=m_2=10^7\,\msun$ 
needs a longer time to merge than a binary with $m_1=10^7\,\msun\gg m_2$, 
as it must eject a larger number of stars. 
A comparison with a straightforward extrapolation of Milosavljevic \& Merritt (2001) N-body 
results shows that the scheme we adopt tends to overestimate the binary evolution timescale
by about a factor 3 (scaling to real galaxies like M32 and M87). Two main factors contribute to 
this 
discrepancy: first, Milosavljevic \& Merritt (2001) let the slope of the density profile change 
smoothly during the hardening of the binary, ending with a shallow cusp $\propto r^{-1}$ rather 
than with a flat core; and second, they take into account the Brownian motion of the binary, 
which makes the BHs interact with a larger number of stars in the central region. 
Nevertheless, the coalescence timescales estimated in our model may be too short, 
since it is the total stellar density that is allowed to decrease following
equation (\ref{rc}),  not the density of low angular-momentum stars (i.e. we neglect 
the depopulation of the loss-cone); furthermore our model 
assumes replenishment of the stellar cusp after every major merger. Yu (2002)
has recently studied the merger of 
binary SMBHs assuming the central stellar  profiles observed by Faber et al. (1997)
in a sample of local galaxies, and finds that BH coalescence timescales may in some case 
be longer than the Hubble time. 
\begin{figurehere}
\vspace{+0.5cm}
\centerline{
\psfig{file=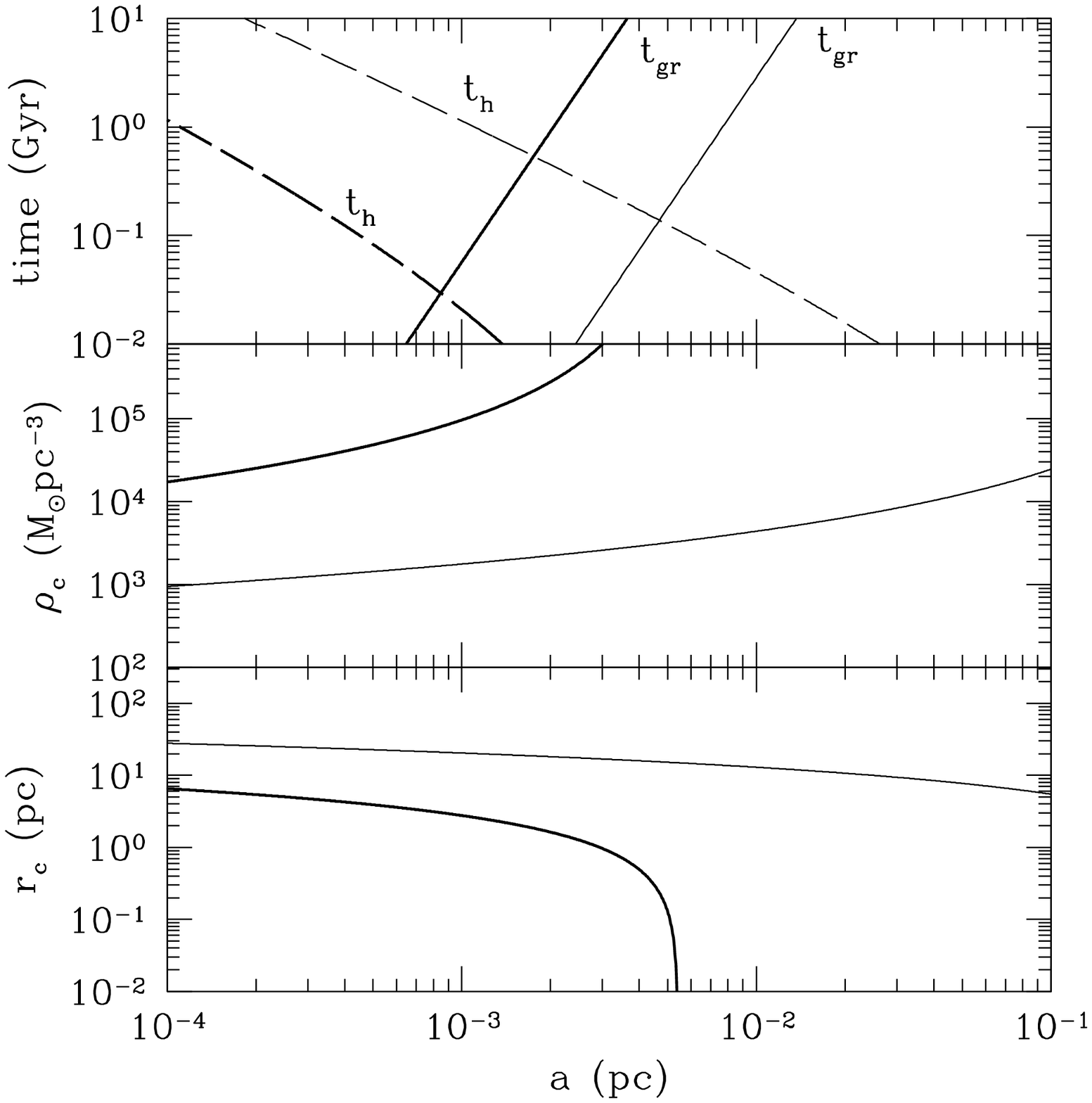,width=2.9in}}
\caption{\footnotesize The evolution (from bottom to top) of core radius, core stellar density,
 and hardening timescale during the shrinking of a SMBH binary in a halo with $\sigma_{\rm DM}=
140\,\kms$. {\it Thin lines:} $m_1=m_2=10^7\,\msun$. The total stellar mass ejected prior 
to coalescence is 5$(m_1+m_2)$. {\it Thick lines:} $m_1=10^7 \msun$, $m_2=10^5 \msun$. 
The total stellar mass ejected prior to coalescence is 2$(m_1+m_2)$.}
\label{fig6}
\end{figurehere}
\vspace{+0.4cm}

\subsection{Triple BH interactions}

The dynamical evolution of SMBH binaries may be disturbed by a third  
incoming BH, if another major merger takes place before the pre-existing binary has 
had time to coalesce (e.g. Hut \& Rees 1992; Xu \& Ostriker 1994). In a minor merger 
the intruder BH 
is stripped of most of the surrounding dark and luminous matter; the ensuing long 
dynamical friction timescale does not allow a close encounter between the central
binary and the intruder. Within our scheme these BHs remain wandering in galaxy halos
through successive mergers. 
If the incoming hole reaches the sphere of influence (determined in our model by the
hardening distance $a_h$) of the central binary, the three BHs are likely to undergo a 
complicated resonance scattering interaction, leading to the final expulsion 
of one of the three bodies (gravitational slingshot). 
Typically an encounter between an intruder of mass $m_\intd$ smaller than both 
binary members leads to a scattering event, where the binary recoils by momentum 
conservation and the incoming lighter BH is ejected from the galaxy nucleus. 
The binary also becomes more tightly bound, each such encounter typically increasing
its binding energy $E_b$ by the amount $\langle \Delta E/ E_b\rangle\approx 0.4\,
m_\intd/(m_1+m_2)$ (Hills \& Fullerton 1980; Colpi, Possenti,
\& Gualandris 2002). By contrast, when the intruder is more massive than one or both  
binary components, the probability of an exchange is extremely high: the incoming hole
becomes the member of a new binary, and the lightest BH of the original pair gets
ejected (Hills \& Fullerton 1980). For mass ratios $m_\intd/(m_1+m_2)\lesssim2$,
most of the increase in the binding energy of the pair is due to the actual shrinking of 
the orbit, while above this value the binding energy rises mainly due to the replacement of 
a low mass member by a more massive BH. In the latter case, and for head-on collisions and
equal mass binaries, the fractional increase in binding energy is approximately
constant with a value of 3.1 (Hills \& Fullerton 1980).
If the binary is hard the kinetic energy and momentum of the intruder are much
lower than the orbital binding energy and the recoil momentum of the ejected body
(we have checked a posteriori that this is true in nearly all cases of triple interactions). 
Conservation of energy and momentum in the interaction allows then to estimate the recoil 
velocity of the binary and intruder. 
Let $m_\ej$ be the mass of the lightest of the three BHs, and 
$m_{\rm bin}$ the mass of the final binary, i.e $m_\ej=m_\intd$ and $m_{\rm bin}=m_1+m_2$ 
in a scattering event, $m_\ej=m_2$ and $m_{\rm bin}=m_\intd+m_1$ for exchanges. The 
kinetic energy of the ejected BH and of the binary after the encounter will then be
\begin{equation}
K_{\rm bin}=\frac{\Delta E}{1+(m_{\rm bin}/m_\ej)}, 
\label{kbin}
\end{equation} 
\begin{equation}
K_\ej=\frac{\Delta E}{1+(m_\ej/m_{\rm bin})}.
\label{kej} 
\end{equation} 
We adopt a simple scheme, where: 
\begin{itemize}

\item if $m_\intd<m_2$ a scattering event occurs, with  
$\langle \Delta E/E_b\rangle=0.4\,m_\intd/(m_1+m_2)$. The new semi-major axis is 
$a_1=a_0/[1+0.4\,m_\intd/(m_1+m_2)]$.

\item if $m_2<m_\intd<2\,(m_1+m_2)$ an exchange with $\langle \Delta E/E_b\rangle=0.4\,
m_\intd/(m_1+m_2)$ takes place.  The new semi-major axis is 
$a_1=a_0(m_\intd/m_2)[1+0.4\,m_\intd/(m_1+m_2)]$.

\item if $m_\intd>2\,(m_1+m_2)$ again an exchange happens, with  
$\langle\Delta E/E_b\rangle=0.9$ (we have rescaled the value of 3.1 given by Hills \& 
Fullerton 1980 for head-on collisions to account for a distribution of impact parameters).
The new semi-major axis is $a_1=0.53\,a_0 (m_\intd/m_2)$.
\end{itemize}

In all cases we have used the equations (\ref{kbin}) and (\ref{kej}) to estimate the 
recoil velocities. All triple interactions are followed along the merger tree, as they
modify the binary separation during each encounter. At high redshift 
we find that the increase in binding energy causes the binary to shrink to a 
separation small enough that coalescence by gravitational radiation occurs, 
since most of these encounters happen among approximately equal-mass systems.  
At later epochs events with low $m_\intd/(m_1+m_2)$ mass ratios are more common 
(see Figure \ref{fig7}), and the binding energy increases only slightly after 
such interactions.  

\begin{figurehere}
\vspace{+0.5cm}
\centerline{
\psfig{file=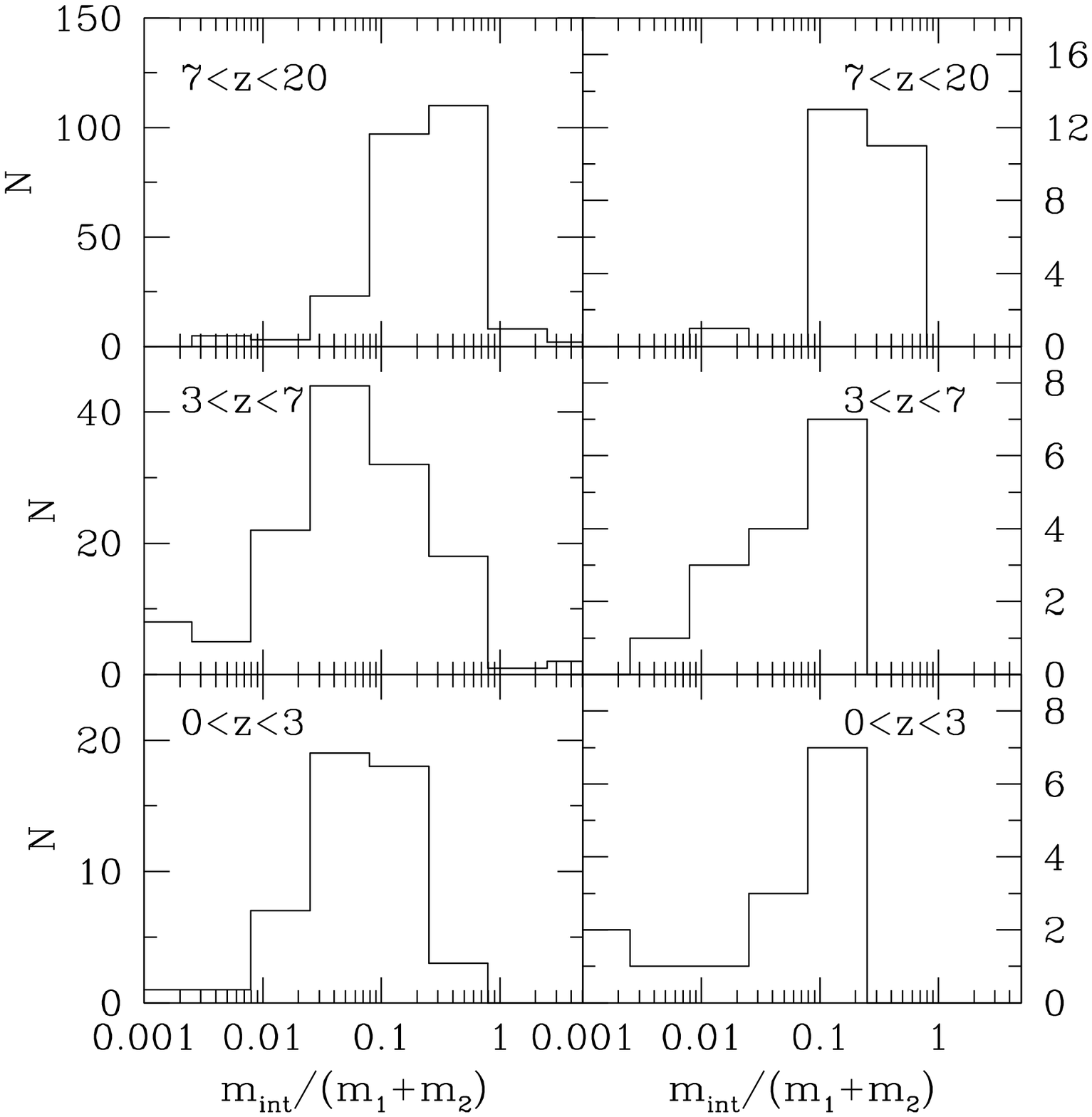,width=2.9in}}
\caption{\footnotesize The number of triple interactions for different ratios between  
the intruder BH mass $m_\intd$ and the mass $m_1+m_2$ of the binary, in different redshift
intervals.  {\it Left panel:} $\sigma_{\rm DM}=250\,\kms$. {\it Right panel:} $\sigma_\DM=150\,
\kms$. This histogram includes results from all 20 Monte Carlo realizations of the same halo mass.
At very high redshift ({\it upper panel}, $7<z<20$) equal mass system interactions are 
more common, while at low redshift ({\it lower panel}, $0<z<3$) a high-mass binary typically 
interacts with an intruder of  much lower mass. At intermediate-high redshift 
({\it middle panel}, $3<z<7$) a transition regime occurs.}
\label{fig7}
\end{figurehere}
\vspace{+0.4cm}

What happens to a BH pair$+$intruder system after the slingshot mechanism? 
In a SIS$+$core halo, the gravitational potential is
\begin{equation}
\Phi(r)=-2\sigma^2_\DM \times \left\{
\begin{array}{ll}
{1 \over 2}
+\ln {r_{\rm vir} \over  r_c}-{1 \over 6} 
\left( {r \over r_c}\right)^2 &\mbox{$r<r_c$} ;\\
~&\\
1 +\ln {r_{\rm vir} \over  r} -{2 \over 3} 
{r_c\over r} & \mbox{$r_c<r<r_\vir$} ;\\
~&\\
{r_{\rm vir}  \over r} -{2 \over 3} {r_c \over r} & \mbox{$r>r_\vir$} ,
\end{array}
\right. 
\end{equation}
where the core radius $r_c$ is that created by the hardening of the binary at the time of the 
triple interaction. If the kick velocity of the binary and/or single BH exceeds the escape 
speed $v_{\rm esc}=\sqrt{2|\phi|}$, the hole(s) will leave the galaxy altogether. We find that 
the recoil 
velocity of the single hole is larger than $v_{\rm esc}$ in 99\% of encounters. The binary is 
ejected instead in only 8\% of the encounters (Figure \ref{fig8}) and typically at very high 
redshifts, when all BHs are in the same mass range.

For equal mass holes both the binary and the single BH are ejected from radius $r_{\rm in}$
to infinity when the orbital velocity $V_{\rm bin}$  of the binary satisfies the condition 
\beq
V_{\rm bin}>7.7\sqrt{\phi(r_{\rm in})}.
\eeq
The 1\% of single BHs not escaping 
their host halos are typically slung to the periphery of the galaxy with consequently long 
dynamical friction timescales; most of the binaries recoil instead within the 
core and fall back to the center soon afterwards, with $t_{\rm df}<0.01\,$Gyr.   
Since most of the ejected holes escape their hosts, in our scheme the number of 
{\it wandering} BHs due to the gravitational slingshot and retained within galaxy halos
is thus significantly lower (by about a factor of 50) than that left over by minor mergers. 
\begin{figurehere}
\vspace{+0.5cm}
\centerline{
\psfig{file=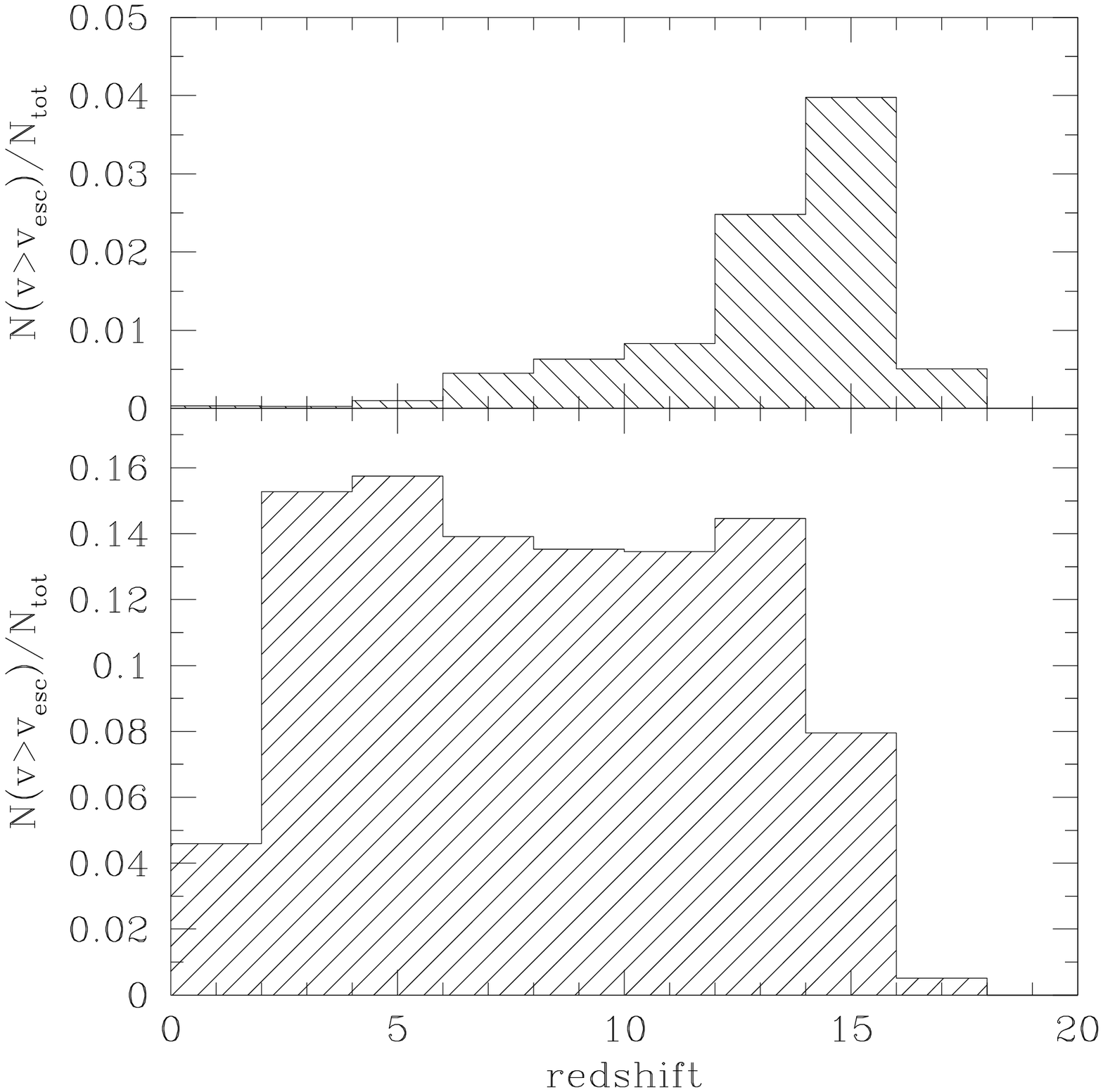,width=2.9in}}
\caption{\footnotesize The fraction of BH binaries ({\it upper panel}) and 
single BHs ({\it lower panel}) with recoil velocity $v$ larger than the escape speed, 
as a function of redshift. The histogram includes results from all Monte Carlo realizations. }
\label{fig8}
\end{figurehere}
\vspace{+0.4cm}
\section{Implications}

In this section we discuss some of the consequences predicted by our fiducial 
scenario (and
a few variants) for the growth of SMBHs in the nuclei of galaxies.

\subsection{The quasars luminosity function}

In our framework quasar activity is triggered by major mergers and SMBHs 
accrete at the Eddington rate, $\dot m_E=4\pi Gm_pm_{\rm BH}/(c\sigma_T\epsilon)$,
where $\epsilon$ is the radiation efficiency.
Accretion starts after about one dynamical timescale and lasts until a mass given by 
equation (\ref{macc_eq}) has been added to the hole. Rest mass is converted to 
radiation with 
a 10\% efficiency; only a fraction $f_{\rm B}=0.08$ of the bolomeric power is radiated in 
the blue band. We have compared theoretical luminosity functions (LF) at different redshifts
with the most recent determination of the quasar blue LF from the 2dF survey ($0.3<z<2.3$, Boyle 
\etal 2000) and the SDSS ($3.3<z<5$, Fan \etal 2001a). The 2dF LF is a double power-law, which
we have extrapolated beyond redshift 2 assuming pure luminosity evolution; the best
fitting parameters for a $\Lambda$CDM cosmology are given by Boyle \etal (2000).  
The SDSS samples only the very bright end of the LF; Fan \etal (2001a) fit a single 
power-law to the data. 

A detailed comparison at very early times required the use of \emph{ad hoc} merger trees
to simulate very luminous quasars in very massive, rare dark halos at high redshifts. 
For this purpose we have simulated the merging histories of $10^{15}\,\msun$ parent halos 
starting from $z=3.5$, and applied the evolutionary scheme for their SMBHs outlined 
in the previous sections. As shown in Figure \ref{fig9}, our simple model reproduces 
reasonably well the observed LF of optically-selected quasars in the redshift 
range $1<z<5$. 

\begin{figurehere}
\vspace{+0.5cm}
\centerline{
\psfig{file=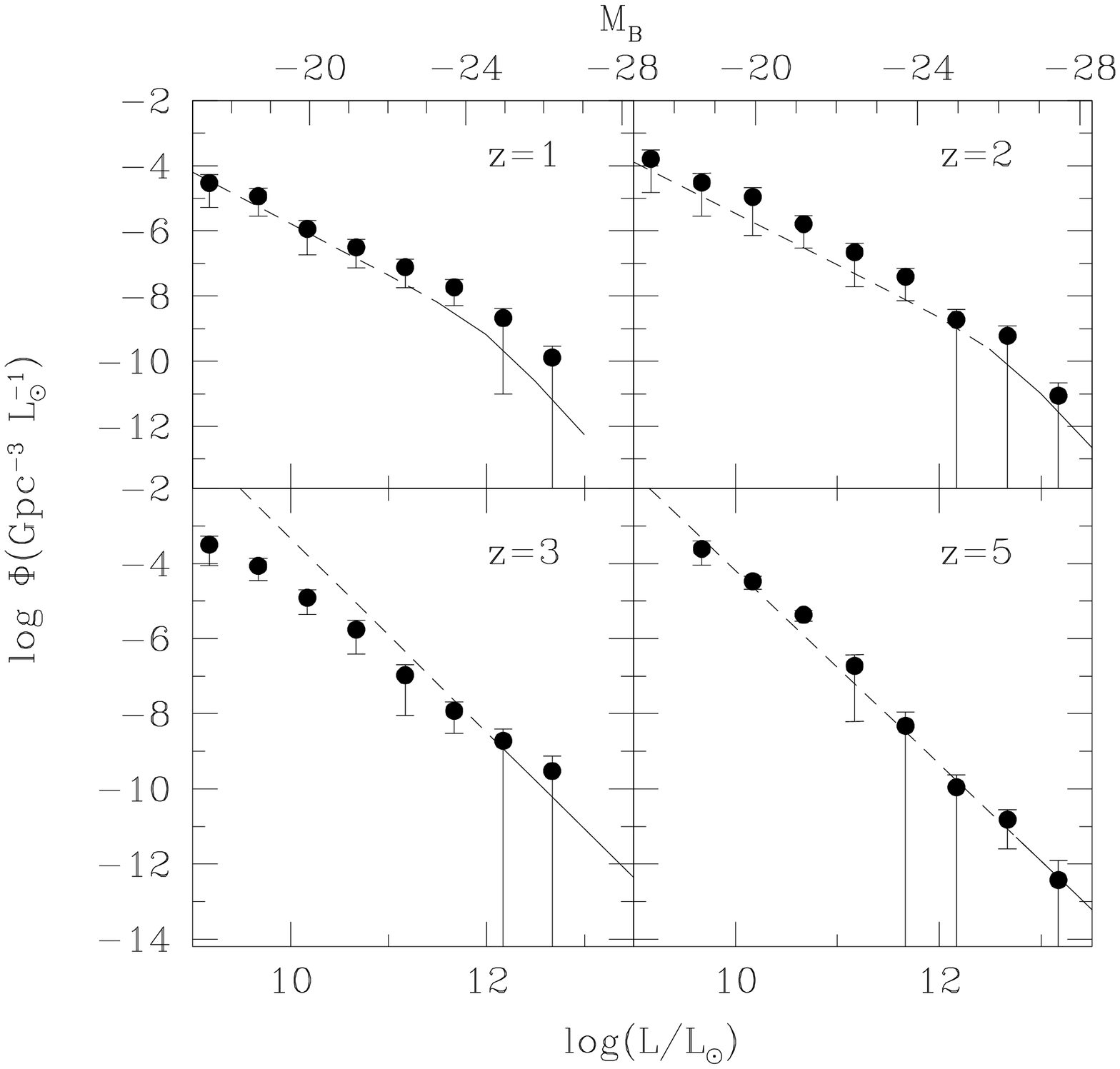,width=2.9in}}
\caption{\footnotesize The B-band luminosity function of quasars at various redshifts.
{\it Filled circles:} the values predicted from our SMBH assembly and merging model history. 
Error bars indicate the poissonian error in the counts.
{\it Solid lines at $z=1,2$:} 2dF LF. {\it Solid lines at $z=3,5$:} SDSS LF. The dashed 
lines show the extrapolation
to faint magnitudes of the best fit LF from Boyle \etal (2000) ($z=1,2$), and Fan 
\etal (2001a) ($z=3,5$).
}
\label{fig9}
\end{figurehere}
\vspace{+0.4cm}

The slope at low luminosities matches the one inferred by Boyle \etal, 
and is considerably flatter than the extrapolation of the SDSS power-law.  
The tendency to overestimate the number of bright QSOs at $z\lta 1$ may be less 
severe due to the presence of a substantial population of (optically) obscured luminous 
AGNs at low redshifts suggested by recent {\it Chandra} results (e.g. Rosati \etal 2002;
Barger \etal 2001), which may help reducing the discrepancy.

To assess the impact on our results of making seed BHs 
more common or rarer, we have run realizations that place $m_\bullet=150\,\msun$ 
BHs at $z=20$ in 3-$\sigma$ (lower bias) and 4-$\sigma$ (higher bias) peaks instead of 
the fiducial 3.5-$\sigma$. We find that we can still reproduce the observed quasar 
LF luminosity function by changing the major merger threshold, from $P>0.3$ in the 
lower bias case, to $P>0.1$ in the higher bias case. 
We have also run a case where the initial, seed BH mass is $m_\bullet=1000\,\msun$ 
instead of 150 $\msun$. We find little change at $z<5$. The number of triple interactions 
at $z>5$, however, increases; this is due to the fact that the hardening timescales 
are now longer, as more massive holes create larger core radii (see eqs. \ref{rc}, \ref{hardt}). 
Overall, the number of BHs ejected in the intergalactic medium (IGM) goes up by a factor of 2.

\subsection{The $m_{\rm BH}-\sigma_c$ relation}

In Figure \ref{fig10} the local $m_{\rm BH}-\sigma_c$ relation predicted by our scheme 
for the assembly history of SMBHs is compared to the fit given by Ferrarese (2002).

The three curves in the lower corner show the $m_{\rm BH}-\sigma_c$ 
relation we would obtain assuming that when halos merge their
BHs coalesce immediately, their masses sum directly, and there is no gas accretion. 
The slope is flatter and the normalization lower than observed. The steepening of the 
relation for initial seed BHs in higher $\sigma$ peaks is a natural outcome 
of biasing. Assuming the entropy-area relation for the merging BHs tends instead to 
flatten the curves. The incapacity of mergers alone to yield the observed 
$m_{\rm BH}-\sigma_c$ has been pointed out by Ciotti \& van Albada (2001) using 
an argument based on the Fundamental Plane.

\begin{figurehere}
\vspace{+0.5cm}
\centerline{
\psfig{file=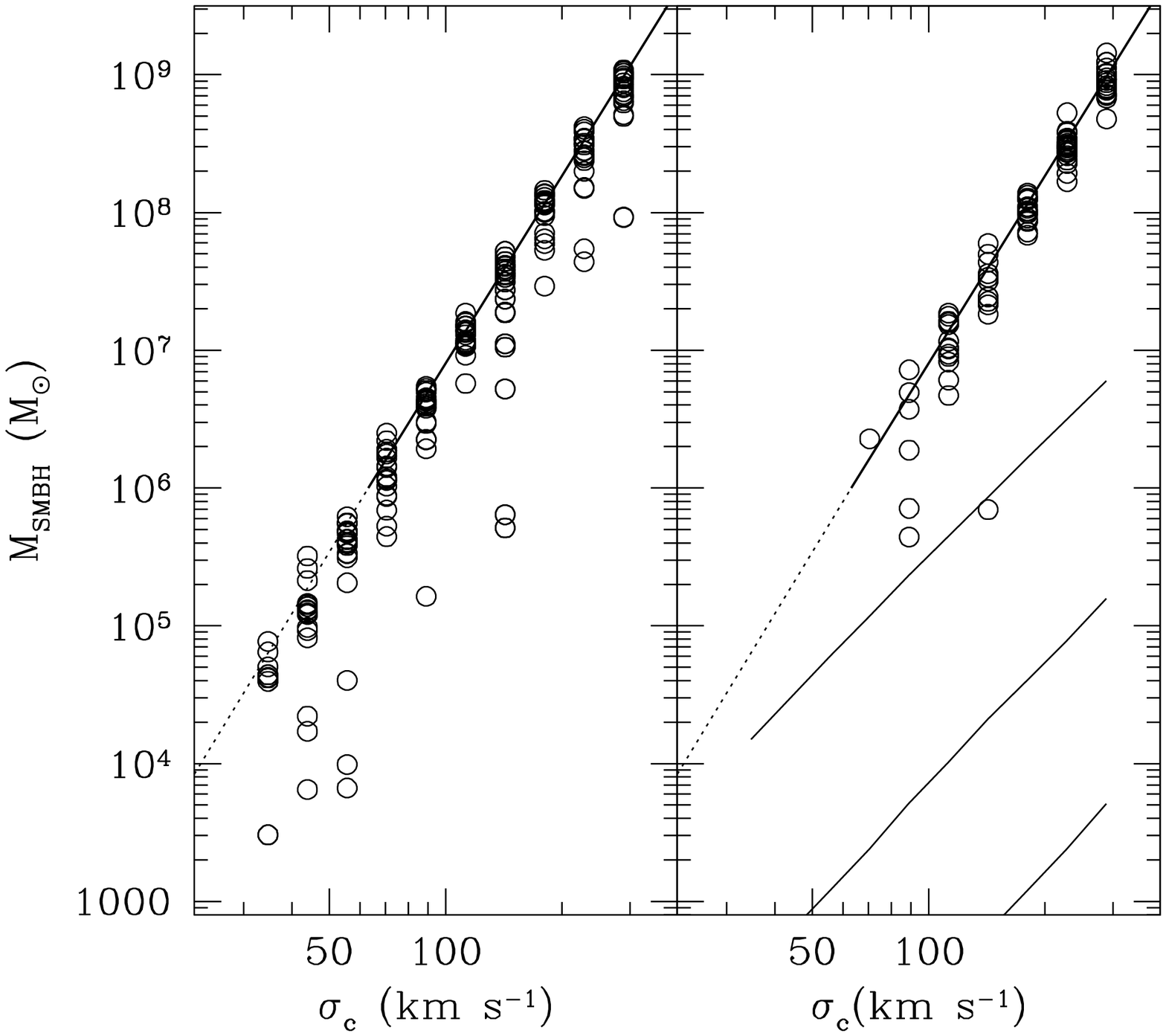,width=2.9in}}
\caption{\footnotesize 
The $m_{\rm BH}-\sigma_c$ relation at $z=0$. Every circle represents one nuclear 
BH in a halo of given $\sigma_c$. We started at z=0 with a discrete grid of halo masses 
(hence, with a discrete grid of $\sigma_c$, see section 3.2 and 3.3), and 
performed several simulations for each mass.
{\it Left panel}: the circles mark the 
results for our fiducial model with seed BHs in 3.5-$\sigma$ density peaks. The solid line
shows Ferrarese's (2002) best fit; its extrapolation to low $\sigma_c$ values is depicted
by a dotted line.   
The holes deviating from the relation are hosted in 
galaxies that experienced their last major merger at $z>1.5$. Since then their host halos
have grown due to minor mergers. {\it Right panel}: same for seed BHs in 4-$\sigma$ 
peaks. The three curves below (from top to bottom, initial seed BHs in 3, 3.5, and 
4-$\sigma$ peaks) show the $m_{\rm 
BH}-\sigma_c$ relation obtained assuming that when halos merge their
BHs coalesce immediately, and there is no gas accretion. Note how the slope is 
flatter and the normalization lower than observed. 
}
\label{fig10}
\end{figurehere}
\vspace{+0.4cm}

The scatter in our model largely reflects the time elapsed since the last major merger. 
While the final SMBH mass is set by the last episode of gas accretion and BH-BH coalescence, 
the host halo mass keeps growing through minor mergers and the accreting of small dark matter
clumps with $M<M_{\rm res}$.  We find that lowering the threshold for major mergers has the 
effect of 
smoothing the mass assembly history of SMBHs, tightening the $m_{\rm BH}-\sigma_c$ relation.  
Figure \ref{fig11} illustrates the fact that the mass-growth history of SMBHs does not 
totally reflect that of their host halos and follows a more complex pattern, with a number
of rapid accretion episodes. 

The assembly history of two SMBHs is shown in Figure \ref{fig12};
one BH (labeled as `1') ends in a main, massive halo at the present epoch, and the other 
(`2') in a satellite at $z=2.3$. 
The final mass of the two BHs is mostly due to gas 
accretion and does not depend on the initial conditions, i.e., whether the seed holes are 
hosted in the 3 or 3.5-$\sigma$ peaks. The number of major mergers between halos hosting SMBHs
is larger in the 3-$\sigma$ peak case; only a fraction ends with the coalescence of the two 
holes before a triple BH interaction takes place, as gravitational slingshots are now 
more common. As an example, for a 
Milky Way-sized halo, the fraction of major mergers that ends up with the coalescence of the
binary without a slingshot drops from 75\% in the 3.5-$\sigma$ fiducial case to 
30\% for $\sigma=3$. 
\begin{figurehere}
\vspace{+0.5cm}
\centerline{
\psfig{file=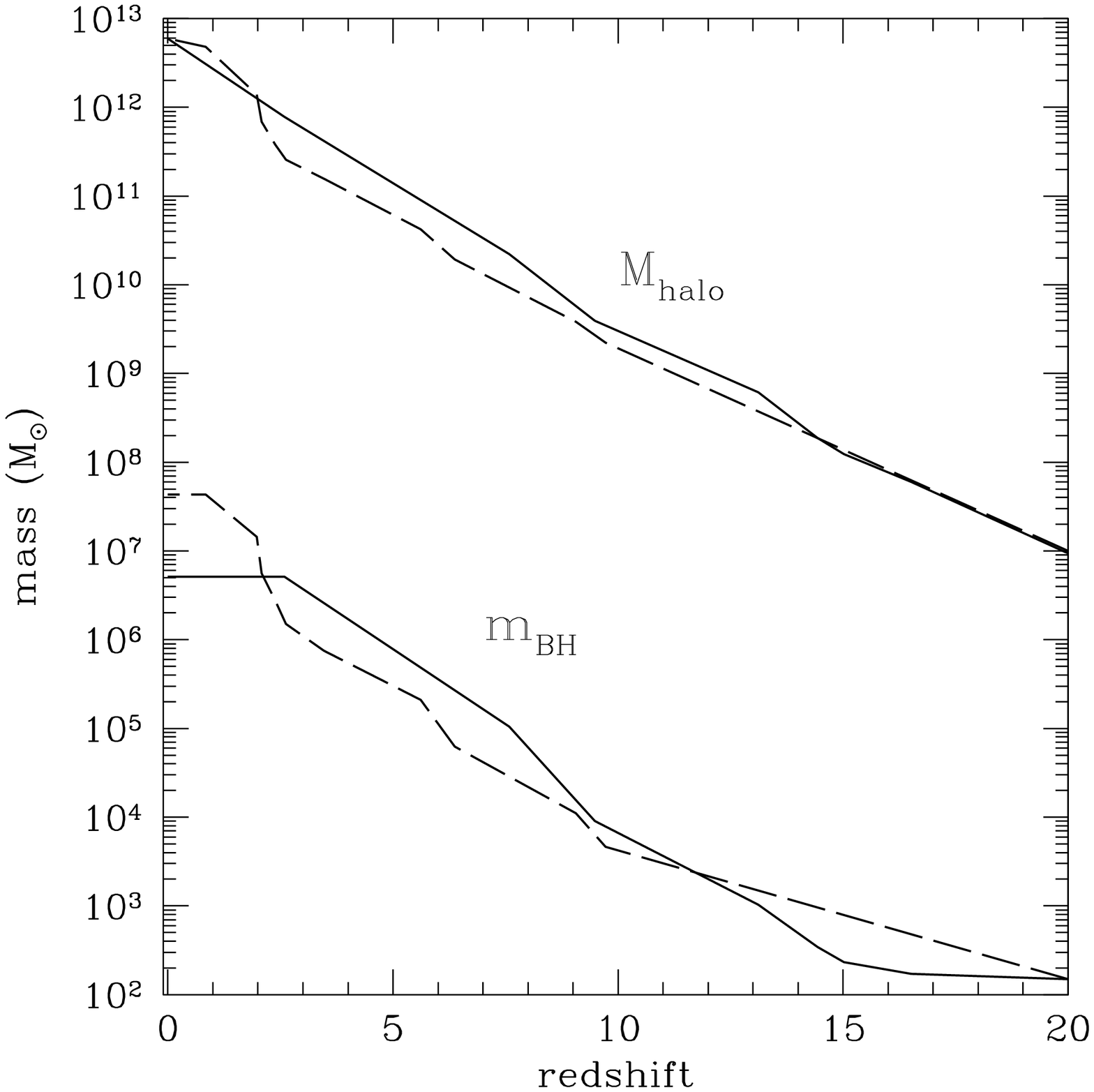,width=2.9in}}
\caption{\footnotesize Two different realizations for the mass-assembly history of a 
galaxy halo with today velocity dispersion $\sigma_\DM=160\,\kms$ and its central SMBH. 
{\it Dashed line:} the halo experiences its last major merger at low redshift, and its hole
follows the observed $m_{\rm BH}-\sigma_c$ relation. {\it Solid line:} the halo has its final 
major merger at $z=2.6$, when only a small fraction of its dark mass was already in place.
The mass of its SMBH is only about 10\% of that expected from the 
$m_{\rm BH}-\sigma_c$ relation.}
\label{fig11}
\end{figurehere}
\vspace{+0.4cm}

\begin{figurehere}
\vspace{+0.5cm}
\centerline{
\psfig{file=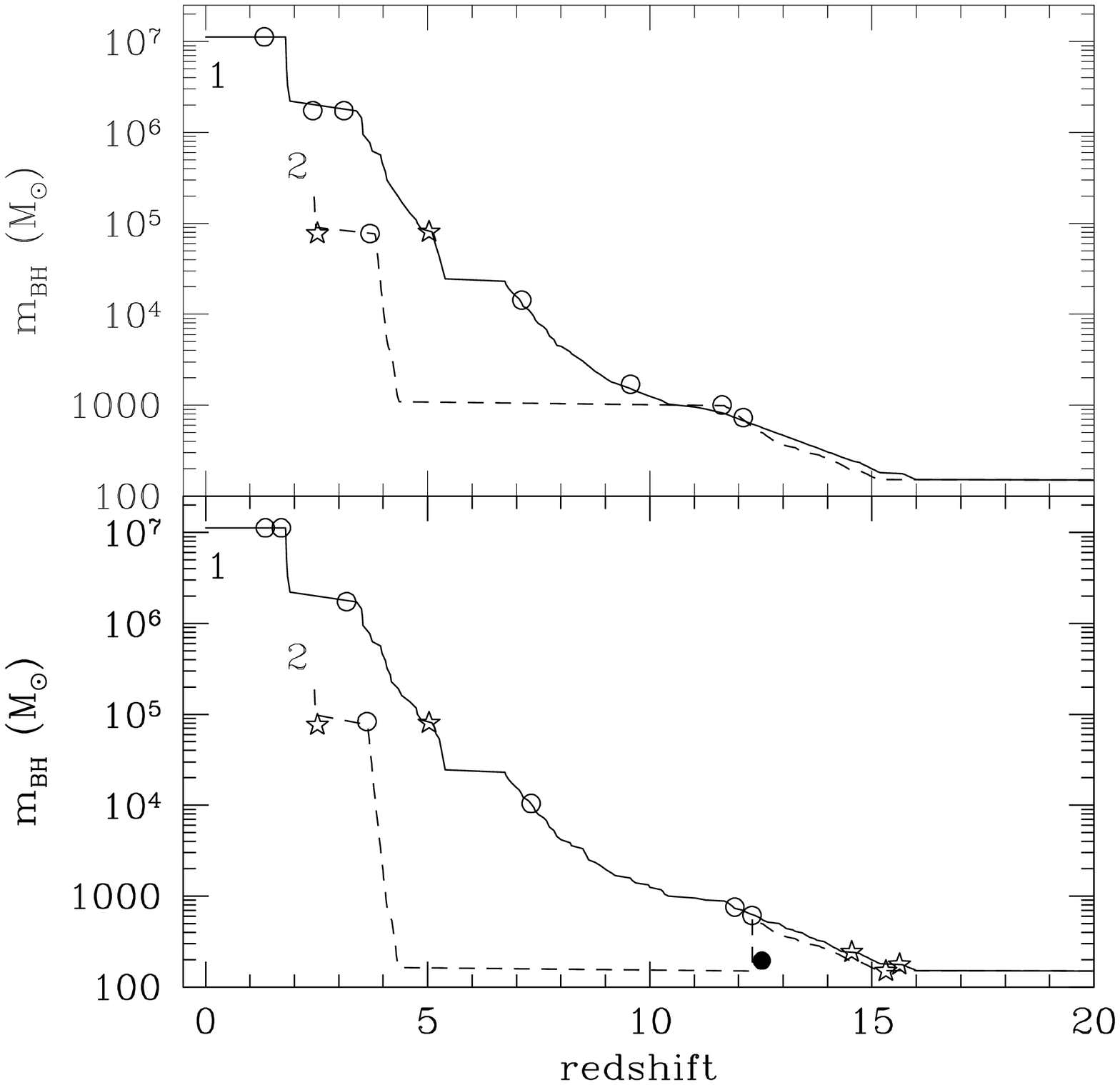,width=2.9in}}
\caption{\footnotesize Mass-growth history of two SMBHs, one ending in a massive 
halo (`1') with $\sigma_\DM=185\,\kms$ at $z=0$, and one in a satellite (`2')
with $\sigma_\DM=80\,\kms$ at $z=2.3$. 
{\it Upper panel}: seed holes in 3.5-$\sigma$ peaks at $z=20$. 
The BH mass grows after every major merger event due to gas accretion, independently of
whether the other merging galaxy hosts another SMBH or not. The starred symbols mark the 
redshift of the major mergers of two parent halos both hosting a SMBH, but not ending up 
with the coalescence of the binary. The circles mark the redshift when two SMBHs coalesce.  
Note how most of the mass of the lighter hole is gained in the most recent accretion episodes. 
{\it Lower panel:} same but for seed BHs in 3-$\sigma$ peaks. The 
number of mergers not ending up with the coalescence of the two SMBHs (indicated by stars) 
is larger compared to the higher bias case. The black dot marks the rare event (occurring in the 
satellite galaxy) of a {\it decrease} of the central BH mass, caused by the ejection of 
the binary in a triple BH interaction.}
\label{fig12}
\end{figurehere}

\subsection{Wandering BHs}

It appears inevitable that significant numbers of triple BH interactions will take place 
at early times if the formation route for the assembly of SMBHs goes back to the very first 
generation of stars, and even more so if the seed holes are more numerous and populate the
low-$\sigma$ peaks. As discussed in \S\,3, our scheme predicts, along nuclear SMBHs 
hosted in galaxy bulges, a number of wandering BHs that are largely 
the result of minor mergers rather than of low energy slingshots. In practice, for minor 
mergers, the dynamical friction timescale is longer than the Hubble time, and at $z=0$ 
the BHs are still on their way to the galactic center. Moreover, rare high energy slingshots 
can eject the lighter holes out of the halo, giving origin to a population of 
free-floating ``intergalactic BHs''. Even rarer events occur when the lighter BH 
and the binary are both ejected into the intergalactic medium.  

The total mass in wandering BHs ranges from 1\% to 10\% of the mass of the central SMBH, for 
halos with $\sigma_\DM=50\,\kms$ and $\sigma_\DM=300\,\kms$, respectively. 
Figure \ref{fig13} shows the mass function of wandering BHs at $z=0$ for two 
typical halos, $\sigma_\DM=100 \kms$ and $\sigma_\DM=200 \kms$.

\begin{figurehere}
\vspace{+0.5cm}
\centerline{
\psfig{file=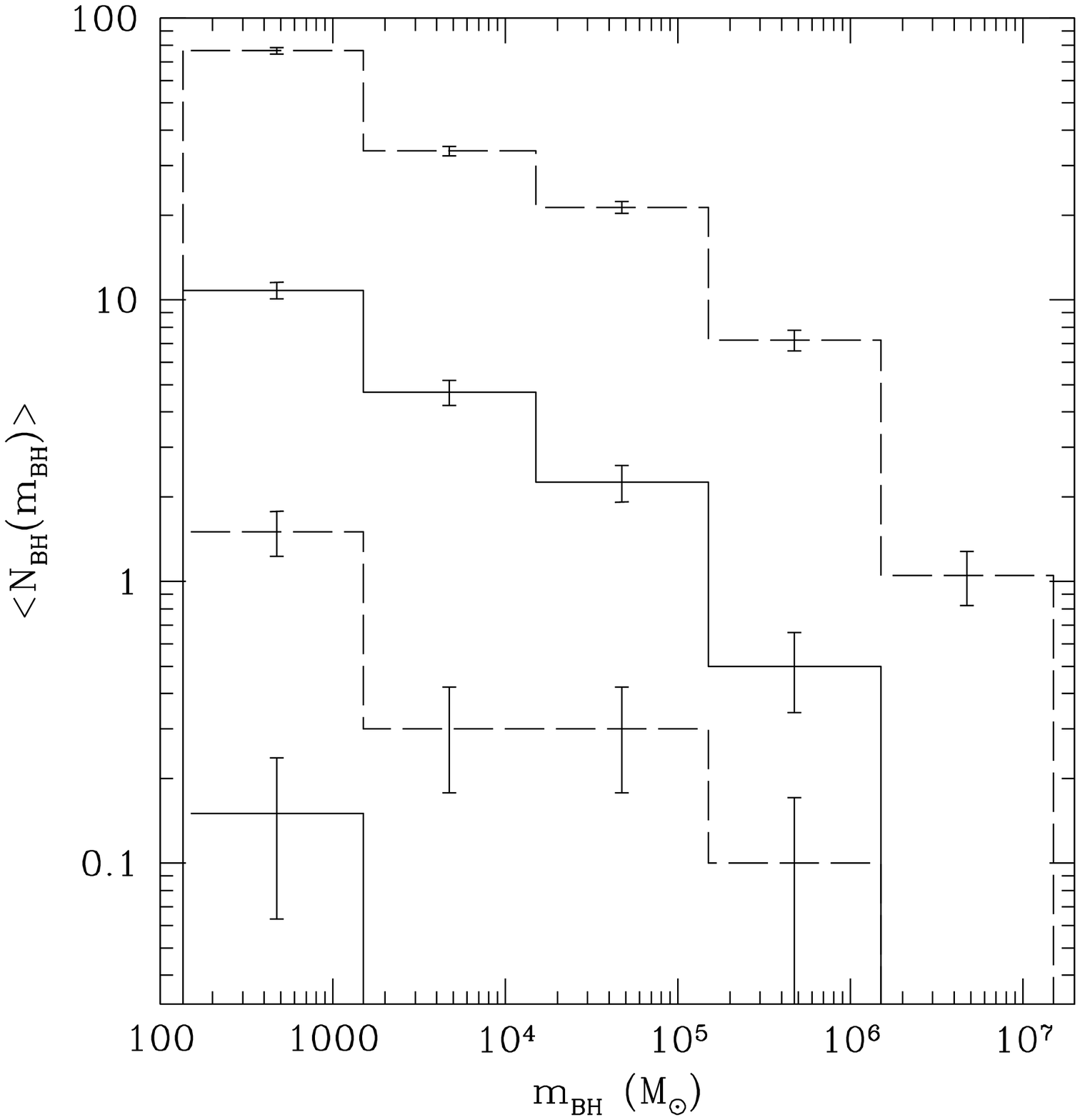,width=2.9in}}
\caption{\footnotesize Mass function of wandering BHs at $z=0$, averaged over 
20 Monte Carlo realizations of a galaxy-sized halo ($\sigma_\DM=100 \kms$, {\it solid lines}), 
and 20 realizations of a more massive, $\sigma_\DM=200 \kms$ halo ({\it dashed lines}). 
The lower-left histograms give the contribution of slingshots to the mass function in the 
two cases considered.  Errorbars are 1-$\sigma$ Poissonian noise. The nuclear SMBHs 
in these halos have masses $m_{\rm BH}=3.8\pm 0.1\times 10^6\,\msun$ and 
$m_{\rm BH}=1.0\pm 0.1\times 10^8\,\msun$, respectively.
}
\label{fig13}
\end{figurehere}
\vspace{+0.4cm}

In Figure \ref{fig14} we show the total mass in BHs predicted by our scheme, and the 
relative contribution of nuclear, wandering and intergalactic BHs, as a function of 
redshift. At $z=0$ the mass in nuclear BHs is $\simeq 3.5\times 10^5 \msun$ Mpc$^{-3}$, 
within 30\% to the value given by Merritt \& Ferrarese (2001), somewhat larger than the value 
estimated by Yu \& Tremaine (2002) from SDSS data and similar to an early 
 estimate by Salucci \etal (1999). The 
total mass is dominated by nuclear BHs at every epochs, though at low redshifts wandering 
BHs become increasingly more important. 

\begin{figurehere}
\vspace{+0.5cm}
\centerline{
\psfig{file=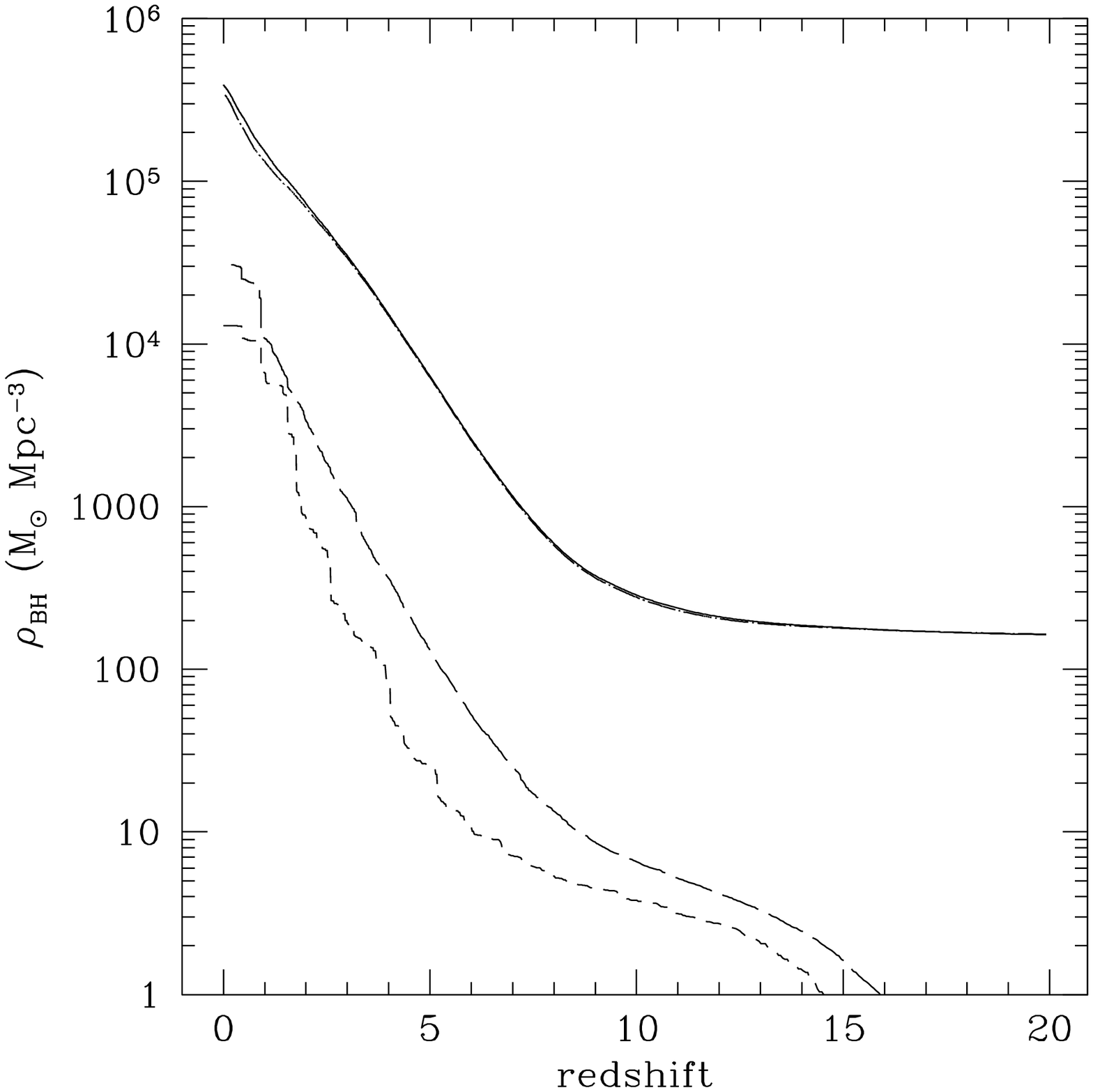,width=2.9in}}
\caption{\footnotesize Contribution of nuclear, wandering and intergalactic BHs to the mass 
density in BHs as a function of redshift. {\it Solid line:} total mass density in all species. 
{\it Dot-dashed line:} nuclear BHs. {\it Long-dashed line:} wandering BHs retained in galaxy 
halos, most of them due to minor mergers. {\it Short-dashed line:} single and binary BHs 
ejected in the IGM after a high energy slingshot event.}
\label{fig14}
\end{figurehere}
\vspace{+0.4cm}

The cumulative local mass function of BHs is displayed in Figure \ref{fig15}. 
The heaviest BH hosted in a given DM halo has experienced several accretion episodes, 
lies in the nucleus, and has a well defined mass according to the tight 
$m_{\rm BH}$-$\sigma_c$ relation. Though the total mass in BHs is dominated, in every
galaxy-sized halo, by the central hole, the same halo hosts a multiplicity of lighter 
wandering BHs, with masses ranging from $150\,\msun$ to approximately one tenth of the mass of 
the nuclear SMBH. Hence the total BH mass density is dominated by wandering holes for masses 
$\lta 10^5 \msun$, and by nuclear ones above.    

Note that on group and cluster scale our scheme would define ``wandering'' all the nuclear 
BHs hosted in satellite galaxies, with only the BH of the central galaxy defined as ``nuclear''. 
On the other hand, on galaxy scales, the cumulative mass contribution of wandering BHs 
is $\sim 10\%$ of the total (wandering+nuclear) mass in a given halo (Figure \ref {fig13}). 
We extrapolate this fraction to larger scales, assuming that 10/11 of the wandering BH mass 
in groups and clusters is associated with nuclear BHs in 
satellite galaxies, and only 1/11 is due to true wandering holes. 
We have then readjusted the mass of wandering and nuclear BHs in groups and clusters 
accordingly.

We keep track of the positions of every single wandering hole within the host halos as it
sinks to the center due to dynamical friction. The positions at the present epoch are 
typically spread between 0.1 and 0.6 of the virial radius. The large population of wandering 
BHs predicted by our model could be associated to the off-center, ultraluminous X-ray 
sources observed in nearby galaxies (e.g., Colbert \& Mushotzky 1999; Makishima et al. 2000; 
Kaaret et al. 2001). 

\subsection{Binary SMBHs and quasars}

The fraction of halos hosting nuclear BHs is shown in Figure \ref{fig16}. 
When computed over all branches of the merger trees, i.e. considering at
$z=0$ only halos with $M>10^{11}\msun$ with smaller and smaller halos 
appearing at higher redshifts, the occupation fraction depends on 
the assumed value of the resolution mass $M_{\rm res}$. To avoid this 
problem, we compute the occupation fraction above a given minimum halo mass. 
\begin{figurehere}
\vspace{+0.5cm}
\centerline{
\psfig{file=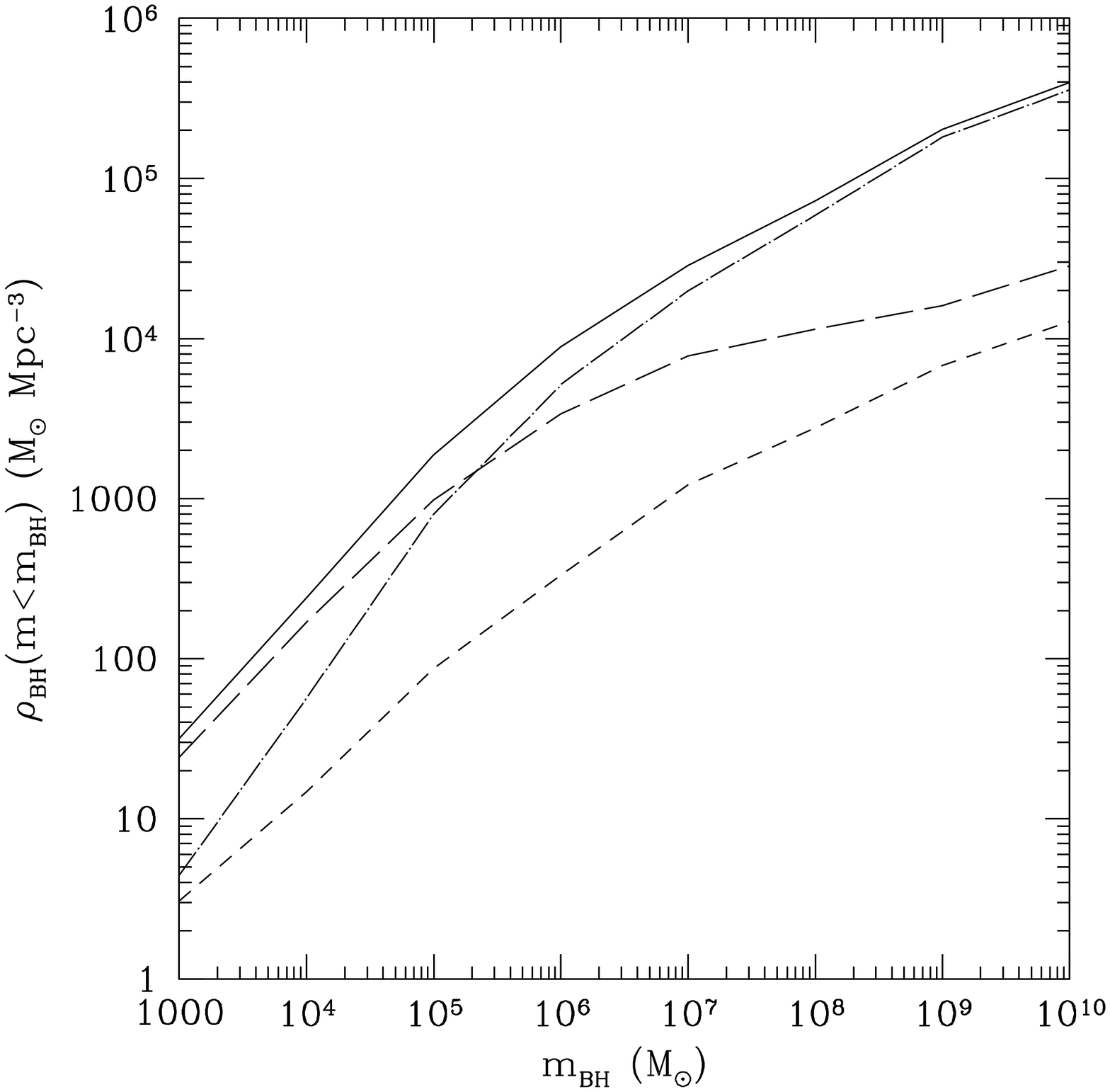,width=2.9in}}
\caption{\footnotesize Cumulative mass density in BHs at redshift $z=0$. The total density is 
dominated  by the nuclear BHs contribution, though the contribute of wandering BHs takes over  
for masses $\lta 10^5 \msun$. {\it Solid line:} total mass 
density. {\it Dot-dashed line:} nuclear BHs. {\it Long-dashed line:} wandering BHs. 
{\it Short-dashed line:} intergalactic BHs.}
\label{fig15}
\end{figurehere}
\vspace{+0.4cm}

As discussed by Menou \etal (2001), this fraction 
initially decreases with cosmic time as lower mass halos lacking nuclear 
BHs become more massive than the assumed threshold by successive mergers.  
Eventually, the occupation fraction starts to increase as the total number of
individual halos drops. Note that our assumed `BH bias' (seed holes in 
$3.5\sigma$ peaks at $z=20$) assures
that all halos more massive that $10^{11}\msun$ actually host a BH at all 
epochs (triple interactions actually slightly reduce the occupation fraction 
somewhat below unity).
In Figure \ref{fig16} we also plot the fraction of all halos containing a 
nuclear BH that actually harbors a binary system. Summed all over the branches
of the tree, this fraction is $\simeq 5-10\%$ today. Almost all massive halos 
at early epochs host a binary, and eventually the fraction of binaries
decreases with cosmic time because of coalescence.

A fraction $\approx 60\%$ of SMBH binaries at $z=0$ has separation larger than 0.1 kpc and 
are still `soft', while $\approx 10\%$ is in an advanced stage of hardening ($a<10$ pc). 
The fraction of binaries strongly depends upon the total coalescence timescale, which, as 
discussed in \S\,3.4, is rather uncertain. We have then run a set of simulations with a 
stellar density artificially decreased by a factor of 2, which implies a 
slower hardening rate by a factor of $\sim 5$, on average. In this case triple 
interactions are more common, due to the increased probability of having a binary 
still in the process of shrinking when a third intruder BH comes along. As a consequence, at 
low redshift ($z\lta 0.1$), the fraction of binaries becomes as large as $20\%$. The increased 
number of binaries is formed by close systems ($a<10$ pc), hosted in galaxies that have 
not experienced any recent major merger. 

\begin{figurehere}
\vspace{+0.5cm}
\centerline{
\psfig{file=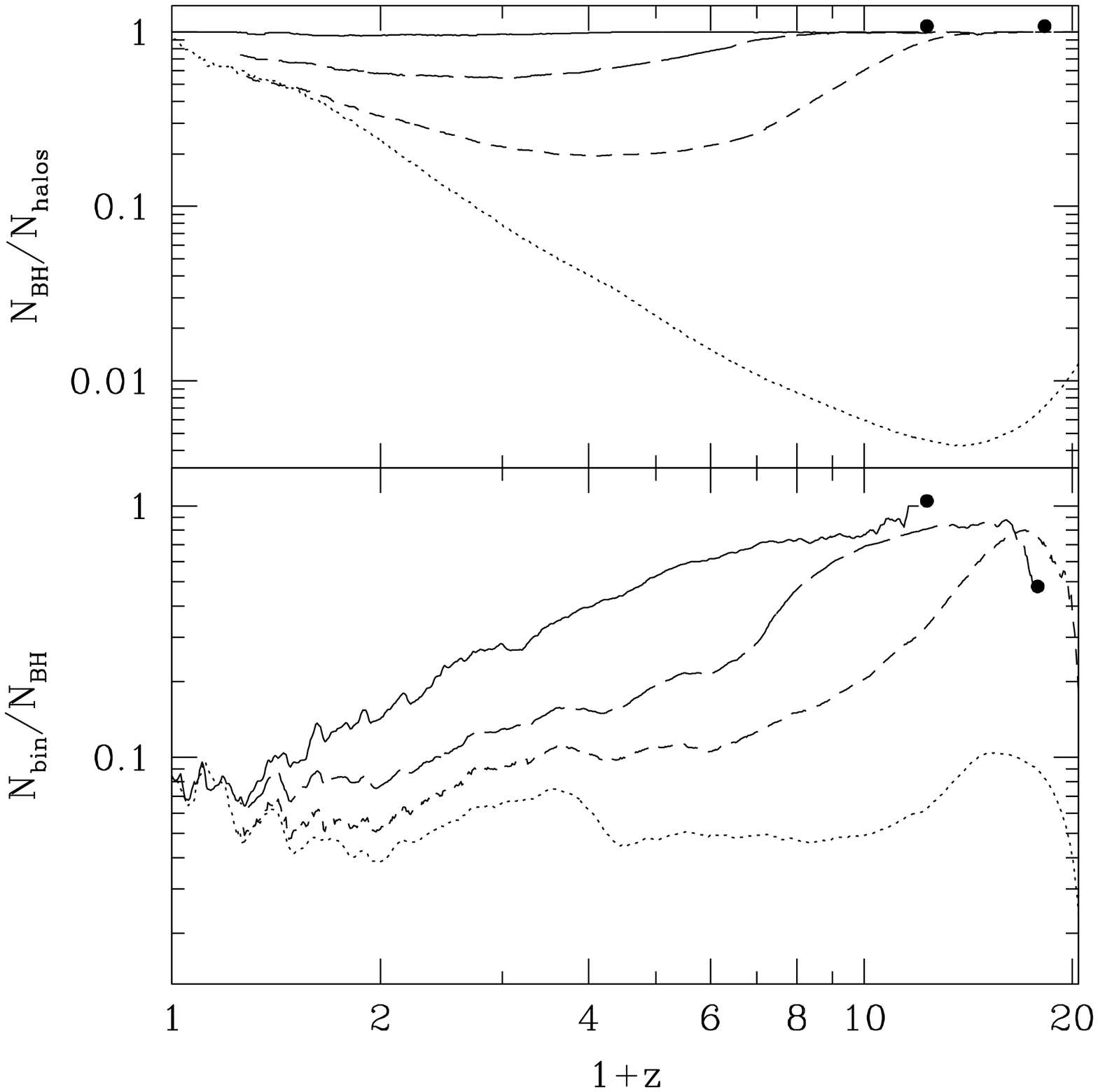,width=2.9in}}
\caption{\footnotesize Fraction of halos hosting at least a nuclear SMBH 
vs redshift ({\it upper panel}) computed weighting over all branches of the 
merger trees ({\it dotted curve}) and imposing different thresholds to 
the halo mass: $M>10^{11}\msun$ ({\it solid curve}), $M>10^{10}\msun$ 
({\it long-dashed curve}) and $M>10^{9}\msun$ ({\it short-dashed curve}). The 
last two quantities are not plotted at $z<0.25$, when the least massive 
halos in the merger trees are already above the adopted threshold.
The mean SMBH occupation fraction is unity at the present epoch, and, when 
weighted over all branches of the merger trees, drops below 
10\ at $z\ge 2$. Black dots mark the epochs when sufficiently massive 
halos appear in the merger trees ($z\simeq 11$ for $M>10^{11}\msun$, 
$z\simeq 16$ for $M>10^{10}\msun$). Approximately 5-10\% of all nuclear BHs are
in binaries ({\it lower panel}). This fraction increases with redshift 
in massive halos.}
\label{fig16}
\end{figurehere}
\vspace{+0.4cm}

Binary quasars are an intrinsically rare phenomenon, as both SMBHs must be active at the same 
time. Observationally, 16 pairs are known in a sample of $\sim 10^4$ QSOs, and among these 16
the confirmed physical associations are less than 10 (Kochanek, Falco, \& Mu{\~ n}oz 1999; 
Mortlock, Webster, \& Francis 1999; Junkkarinen \etal 2001). 
In our scheme, since only the SMBH hosted in the larger halo accretes and radiates
during a major merger, at least two subsequent major mergers are required to give origin to 
a binary QSO system, the satellite BH being activated in the first merger event. 
Note that the mean timescale over which BHs accrete and radiate is $\sim 4\times 10^7$ years,
much shorter than the typical time between two major mergers (Somerville et al. 2001).
In our simulations we find a 
fraction of binary quasars with $L>0.01L_*$ that is $\simeq 1-3\times 10^{-3}$ at $z<4$. 
A similar fraction is found taking a luminosity threshold of $0.1L_*$ ($L_*$ was calculated 
at every redshift from the polynomial evolution given by Boyle \etal 2000).
If instead both BHs involved in a major merger were assumed to accrete, the fraction of 
binary quasars would be two orders of magnitude larger, in conflict with the existing 
data. 

\vspace{+0.5cm}
\section{Summary}
Motivated by the recent discovery of luminous quasars around redshift $z\approx 6$ -- 
suggesting a very early assembly epoch -- and by numerical simulations of the fragmentation 
of primordial molecular clouds in cold dark matter cosmogonies, we have assessed a model 
for the growth of SMBHs in the nuclei of luminous galaxies out of accreting Pop III
seed holes of intermediate masses, the endproduct of the first 
generation of stars in (mini)halos 
collapsing at $z\sim 20$ from high-$\sigma$ density fluctuations. As these pregalactic 
BHs become incorporated through a series of mergers into larger and larger 
halos, they sink to the center owing to dynamical friction, accrete a fraction of the
gas in the merger remnant to become supermassive, form a binary system, and eventually
coalesce. We have followed the merger history of dark matter halos and associated BHs 
through cosmological Monte Carlo realizations
of the merger hierarchy from early times until the present in a $\Lambda$CDM 
cosmology. In our scheme the current mass of SMBHs lurking at the center of galaxy 
accumulates mainly via gas accretion, with BH-BH mergers playing only a secondary role.

The main results of our investigations can be summarized as follows: 

\begin{itemize}

\item a simple model where quasar activity is driven by major mergers 
and SMBHs 
accrete at the Eddington rate a  mass that scales with the fifth power of the 
circular velocity of the host halo, can reproduce the observed
luminosity function of optically-selected quasars in the redshift range $1<z<5$;    

\item hardening of BH binaries takes place efficiently both 
as a result of cuspy stellar density profiles that are replenished after
every major merger and, to some extent, due to triple BH interactions. 

\item although our seed BHs at $z=20$ are very rare (one in every halo collapsing from 3.5-$\sigma$ density 
peaks), the nuclear SMBH occupation fraction is of order unity at the present epoch. It drops to less 
than 10\% only at $z\ge 2$. Had we placed seed BHs in the 4-$\sigma$ density peaks instead, the occupation 
fraction of nuclear SMBH would be approximately 0.6 today;
 
\item the local fraction of binary SMBHs is of order 10\%, with half of 
these systems having a separation larger than 100 pc. (Surviving binary SMBHs
have mass ratios $0.2\pm 0.1$.) This fraction increases with redshift,
so that almost all massive nuclear BHs at early epochs are in binary systems;

\item at $z<4$, binary quasars represent a fraction $1-3\times 10^{-3}$ of all
AGNs more luminous than $0.1L_*$;

\item the long dynamical friction timescales and BH slingshots create a population of 
BHs wandering in galaxy halos and free-floating in the IGM, and
contributing $\lta 10\%$ to the total BH mass density today. For a Milky Way-sized 
galaxy we estimate $\simeq 10$ wandering BHs with mass between 150 and 1000 $\msun$, 
and $\sim 1$ wandering SMBH with $10^5<m_{\rm BH}<10^6\,\msun$ (cf. MR). 
For a halo with $\sigma_\DM=200\kms$ the number of wandering BHs is approximately 
10 times larger. 

\end{itemize}

Because of the sensitivity of our calculations on a number of uncertain parameters, 
these predictions should be regarded only as trends. Yet, we believe our results shed new 
light  on models for the assembly of SMBHs that trace  their hierarchical build-up far 
up in the dark halo merger tree. In a subsequent paper we will explore in detail the 
possibility that the damage done to stellar cusps by binary BHs may be {\it cumulative}, 
together with the detectability of wandering BHs in galaxy halos, free-floating holes in
the IGM, and coalescing SMBHs by a low-frequency gravitational wave experiment such as the 
planned {\it Laser Interferometer Space Antenna} ({\it LISA}).

\smallskip
\acknowledgements
We have benefitted from discussions with L. Ciotti, M. Colpi, D. Merritt, 
J. Ostriker, M. Rees, and G. Taffoni. 
Support for this work was provided by NASA through grants NAG5-4236 and NAG5-11513 
(P. M.),  by NSF grant AST-0205738 (P. M.), and by ASI (M. V.). M. V. thanks 
the Department of Astronomy and Astrophysics, UCSC for kind hospitality. 

{}

\end{document}